\documentclass[twocolumn]{aastex631}

\usepackage{amssymb}
\usepackage{amsmath}
\usepackage{comment}
\usepackage[colorlinks=true]{hyperref}
\usepackage{cleveref}

\shorttitle{Principle of Inhomogeneity II}
\shortauthors{Zhang}
\graphicspath{{./}{figures/}}

\begin{document}

\title{The Inhomogeneity Effect II: \\
Rotational and Orbital States Impact Planetary Cooling}

\correspondingauthor{Xi Zhang}
\email{xiz@ucsc.edu}

\author{Xi Zhang}
\affiliation{Department of Earth and Planetary Sciences, University of California Santa Cruz, Santa Cruz, CA 95064, USA}

\begin{abstract}

We generalize the theory of the inhomogeneity effect to enable comparison among different inhomogeneous planets. A metric of inhomogeneity based on the cumulative distribution function is applied to investigate the dependence of planetary cooling on previously overlooked parameters. The mean surface temperature of airless planets increases with rotational rate and surface thermal inertia, which bounds the value in the tidally locked configuration and the equilibrium temperature. Using an analytical model, we demonstrate that the internal heat flux of giant planets exhibits significant spatial variability, primarily emitted from the nightside and high-latitude regions acting as ``radiator fins." Given a horizontally uniform interior temperature in the convective zone, the outgoing internal flux increases up to several folds as the inhomogeneity of the incoming stellar flux increases. The enhancement decreases with increasing heat redistribution through planetary dynamics or rotation. The outgoing internal flux on rapidly rotating planets generally increases with planetary obliquity and orbital eccentricity. The radiative timescale and true anomaly of the vernal equinox also play significant roles. If the radiative timescale is long, the outgoing internal flux shows a slightly decreasing but nonlinear trend with obliquity. Our findings indicate that rotational and orbital states greatly influence the cooling of planets and impact the interior evolution of giant planets, particularly for tidally locked planets and planets with high eccentricity and obliquity (such as Uranus), as well as the spatial and temporal variations of their cooling fluxes.

\end{abstract}



\section{Introduction} \label{introsec}

In our first study (\citealt{zhangInhomogeneityEffectInhomogeneous2023}, hereafter ``Paper I"), we proposed a general principle that inhomogeneities of the surface and atmosphere of a planet cause a greater cooling effect than a homogeneous planet with the same average incident stellar flux and opacity. This is due to the concave dependence of the surface temperature on incident stellar flux and opacity on terrestrial planets and the convex dependence of the internal heat flux on these variables on giant planets. By Jensen's inequality, the expected value of a convex (concave) function is larger (smaller) than the function of the expected value. We validated the principle using an analytical radiative-convective equilibrium (RCE) model. Our results showed that the mean surface temperature on an inhomogeneous terrestrial planet could decrease by more than 20\%, and the internal heat flux on an inhomogeneous giant planet can increase over an order of magnitude compared to homogeneous cases.

 The analysis of opacity inhomogeneity was limited by only adjusting the overall opacity of the atmospheric column in the analytical model, which cannot capture more intricate patterns present in actual atmospheres. On the other hand, the inhomogeneity of incident stellar flux is relatively simpler to evaluate, as the incident stellar flux pattern on a three-dimensional (3D) spherical planet is well-defined. The time variability of the incident stellar flux pattern is also predictable  utilizing planetary obliquity and orbital eccentricity. Planetary processes such as rotation and atmospheric circulation can homogenize the energy flux distribution, and the heat redistribution efficiency depends on properties such as radiative timescale and thermal inertia. Given the rotational and orbital states of a planet and the energy redistribution processes, it is possible to quantify the inhomogeneity effect of the incident stellar flux on the cooling efficiency of the planet. It will be the focus of this paper.

Furthermore, while the previous theory suggests that any planet with an inhomogeneous stellar flux pattern should emit more energy to space than a planet with a uniform flux, it does not provide insight into how planetary cooling depends on the physical processes and parameters that create the inhomogeneity. Jensen's inequality can only compare inhomogeneous cases to their homogeneous counterparts. Intuitively, one might expect a planet to cool faster with a greater degree of inhomogeneity. For example, maybe a tidally-locked planet cools faster than a rapidly-rotating planet given the same total stellar flux. To test this hypothesis, in this study we introduce a metric for inhomogeneity and expands the theory in Paper I to enable comparisons among planets with varying degrees of inhomogeneity.

After we establish the metric, we can apply the generalized theory to understand how rotational and orbital states, as well as planetary processes, influence energy flux distribution and planetary cooling. We can then reveal how planetary cooling depends on fundamental parameters such as rotation rate, obliquity, eccentricity, heat redistribution efficiency, and thermal inertia. Traditionally, these parameters were not considered in the planetary cooling analysis as they are not directly related to radiative processes. However, these parameters control the energy redistribution on a 3D planet, which greatly influences its cooling. We will first show the example of an airless body to validate our theory and demonstrate how the mean surface temperature depends on planetary properties. The majority of the paper will then focus on understanding how the interior cooling of giant planets depends on different rotational and orbital configurations and heat redistribution efficiency in the atmosphere. 

It has been long recognized that the atmospheres of giant planets can significantly regulate heat flow from the interior and influence planetary cooling (e.g., \citealt{hubbardJovianSurfaceCondition1977}). A larger absorption of stellar flux in the atmosphere can cause the radiative-convective boundary (RCB) to move deeper, which in turn suppresses planetary cooling (e.g., \citealt{guillotGiantPlanetsSmall1996}). To date, most planetary evolution theories adopt a one-dimensional (1D) approach and use the mean stellar flux (e.g., \citealt{graboskeStructureEvolutionJupiter1975, guillotGiantPlanetsSmall1996, marleyAtmosphericEvolutionarySpectral1996a, burrowsNongrayTheoryExtrasolar1997, hubbardComparativeEvolutionJupiter1999, fortneyPhaseSeparationGiant2003, fortneyPlanetaryRadiiFive2007, ginzburgExtendedHeatDeposition2016, komacekStructureEvolutionInternally2017,komacekReinflationWarmHot2020,mollousInflationMigratedHot2020}). These frameworks have primarily focused on the average characteristics of atmospheres and their effects on interior cooling. But they have overlooked other factors that influence the distribution of these properties over time and space and their impact on cooling. Our study emphasizes the dependence of interior cooling on fundamental planetary properties, such as rotational and orbital configurations.

An extreme example is the tidally locked gas giants, also known as hot Jupiters, which experience extreme day-night temperature contrasts to enhance interior cooling (e.g., \citealt{guillotEvolution51Pegasus2002, budajDayNightSide2012,spiegelThermalProcessesGoverning2013}). \cite{rauscherINFLUENCEDIFFERENTIALIRRADIATION2014} argued that the RCB on these planets might be longitudinally uniform if the day-night heat redistribution by fast zonal jets. But even in this limit, the latitudinally inhomogeneous RCB could still enhance interior cooling by up to 50\% compared to a globally uniform RCB in their analysis. In principle, the uniformity of the RCB in real atmospheres is determined by heat redistribution, which is controlled by atmospheric radiation and circulation. A full understanding of the effect of inhomogeneous RCB that varies with both latitude and longitude requires 3D atmospheric dynamical simulations.

Planets in high eccentric orbits, such as HD 80606 b, or with high obliquities, such as Uranus, are other interesting examples where the inhomogeneity effect could be important. These planets experience large temporal and spatial variations in incoming stellar flux (e.g., \citealt{rauscherModelsWarmJupiter2017,ohnoAtmospheresNonsynchronizedEccentrictilted2019,  ohnoAtmospheresNonsynchronizedEccentrictilted2019a}). Some evolution models have tracked the orbital evolution and the flux change through planet migration (e.g., \citealt{millerInflatingDeflatingHot2009}), but a detailed examination of how planetary cooling is impacted by eccentricity and obliquity is still needed.

This paper aims to provide a theoretical foundation to understand the dependence of planetary cooling on various planetary parameters. We used the 1D analytical RCE atmospheric model similar to that in Paper I but expanded it to multiple RCE columns on a 3D planet with different rotational and orbital configurations. Previous studies usually prescribed the internal heat flux as an input at the lower model boundary (called $F_{\rm int}$). Here take a different approach. We calculated the internal heat flux by assuming that the atmospheric temperature at the bottom boundary, deep in the convective layer, is efficiently homogenized over an isobar (i.e., constant pressure level). This allows for exploring the internal heat flux and its variations with time and space and the dependence of the globally averaged internal heat flux on different parameters. 

The presence of horizontally uniform temperatures and non-uniformly distributed internal heat fluxes throughout the convective atmospheres seems to be well-supported for giant planets in our Solar System. For example, the thermal infrared observations of Jupiter from Pioneer 10 and 11 (\citealt{ingersollPioneer11Infrared1975}), Voyager missions (e.g., \citealt{hanelAlbedoInternalHeat1981,hanelAlbedoInternalHeat1983,pirragliaMeridionalEnergyBalance1984}), Cassini spacecraft (e.g., \citealt{nixonAbundancesJupiterTrace2010,zhangRadiativeForcingStratosphere2013}), and ground-based studies (e.g., \citealt{fletcherJupiterTemperateBelt2021,geRotationalLightCurves2019,ortonUnexpectedLongtermVariability2023}) have revealed a roughly uniform distribution of tropospheric temperatures and outgoing thermal emission across latitudes. If we define the outgoing internal heat flux as the \textit{local difference} between the outgoing longwave radiation and incoming solar radiation at the top of the atmosphere, the large stellar flux difference between the equator and the poles implies a large latitudinal contrast in the internal heat flux on Jupiter. In the context of exoplanets, simulations conducted by 3D general circulation models for tidally locked gas giants, which are characterized by significant day-night irradiation contrast, also demonstrate a convergence of temperature profiles in the deep atmosphere (e.g., \citealt{komacekEffectInteriorHeat2022}).

We will first introduce the metric of inhomogeneity using probability theory in Section \ref{sec:theory}. In Section \ref{sec:airless}, we explore the dependence of the mean surface temperature on airless planets on different stellar flux patterns, rotational rates, and surface thermal inertia. In Section \ref{sec:inso}, we examine the effect of inhomogeneously distributed stellar flux on giant planets. We first present several typical cases under physical extremes to demonstrate how the outgoing internal heat flux depends on the rotational configuration. The effect of horizontal heat redistribution on planetary cooling is also investigated. In Section \ref{eosec}, the effects of planetary obliquity and orbital eccentricity on the incident stellar flux distribution and planetary cooling flux are explored. In Section \ref{consec}, the main points are summarized, and the implications of this theory on planetary evolution for tidally locked giant planets, planets in high eccentric orbits, and those with high obliquity are discussed.

\section{Generalized Theory of the inhomogeneity effect} \label{sec:theory}

Jensen's inequality compares only the inhomogeneous case with the homogeneous case. To compare different inhomogeneous planets, the central problem is how to measure the inhomogeneity quantitatively. A simple example is the case with only two columns with varying contrast, as demonstrated in Paper I, in which the inhomogeneity is well defined by the contrast (e.g., the difference between the high value and the low value). However, the metric is not straightforward for a statistical ensemble with a range of values. The inhomogeneity cannot be defined as the variance of the distribution in the form of $f(x)=(x-\overline{x})^2$ where $\overline{x}$ denotes the mean value of $x$, because the variance itself is a simple convex function and cannot represent all types of convex functions.

This problem is related to the ``Jensen Gap" in statistics: $\overline {f(X)}-f(\overline {X})$ where $X$ is data. It is a measure of the difference between the expected value of a function applied to a random variable (i.e., $\overline {f(X)}$) and the function applied to the expected value of the random variable (i.e., $f(\overline {X})$). The problem posed is equivalent to asking what statistical properties control the size of the Jensen gap. Unfortunately, the Jensen gap is still an active research frontier in statistics. Some studies suggested that the Jensen gap can be related to the variance of the variable set, but they had to introduce another new parameter that depends on the variable set (e.g., \citealt{liaoSharpeningJensenInequality2019}).

The incident stellar flux pattern on a planet with a spherical shape in an elliptical orbit generally varies smoothly with time and space. Thus we can propose a metric of inhomogeneity that might be sufficient for investigating the inhomogeneity of the incident stellar flux in most situations. In Appendix \ref{app:convex}, we prove two theorems for the generalized theory of the inhomogeneity effect for discrete and continuous variables, respectively. Both theorems share the same mathematical essence and are briefly stated here.

\textit{Theorem I: For two sets of discrete variables with the increasing order, $X = \{x_1, x_2, ..., x_n\}$ and $Y = \{y_1, y_2, ..., y_n\}$, with the same expected value $\overline{X} = \overline{Y}$, we claim that if the differential sequence $x_i - y_i$ has one sign change from negative to positive as $i$ increases from 1 to $n$, i.e., there exists $y_c$ in $Y$ such that $(y_i - y_c)(x_i - y_i) \geq 0$ for all $i$, then the set $X$ has a larger inhomogeneity than $Y$. For any convex function $f(X)$, the expected value of function $\overline {f(x)}$ is larger in a more inhomogeneous case}.

We have also shown in Appendix \ref{app:convex} that the required condition $(y_i-y_c)(x_i-y_i) \ge 0$ implies the variance of $X$ is larger than $Y$ (i.e., $\mathrm{var}(X) \ge \mathrm{var}(Y)$), but is stronger than the variance comparison. We can also translate the metric using the cumulative density functions (CDF) in the probability theory as follows. 

\textit{Theorem II: Suppose there are two probability density functions (PDFs) $g(x)$ and $h(x)$ with the same expected value of $x$. Let $G(x)$ and $H(x)$ be the CDFs of $g(x)$ and $h(x)$, respectively. If the function $G(x)-H(x)$ has one sign change from positive to negative as $x$ increases, i.e., there exists $x_c \in [0,\infty]$ such that $(x-x_c)\left[G(x)-H(x)\right] \le 0$, then the PDF $g(x)$ yields a larger inhomogeneity than $h(x)$. For any convex function $f(X)$, the expected value of function $\overline {f(x)}$ is larger in a more inhomogeneous case}.

In other words, the slope of a CDF can be used as a measure of inhomogeneity. This can be visualized by comparing the CDFs of two data sets on the same graph, with variables on the vertical axis and cumulative probability on the horizontal axis. A flatter slope indicates a lower degree of inhomogeneity. An extreme example is if one data set has all variables with the same value equal to the mean of the other data set. The CDF of the uniform data set will be a horizontal line at the mean value, while the non-uniform data set will have a steeper trend as it passes that line. The steeper slope indicates a higher inhomogeneity. This special case is the situation of Jansen's inequality.

We can apply the generalized theory to the two-column comparison in Appendix E of Paper I. Graphically, the two variables $X = \{x_0-a, x_0+a\}$ and $Y = \{x_0-b, x_0+b\}$ will yield two lines on the CDF graph. The line with the steeper slope (i.e., the set $X$ if $a \ge b$) exhibits a larger contrast and yields a larger $\overline{f}$. 

The theory can generally be applied to most ``nice" distributions, such as single-Gaussian or power-law distributions, but it is not suitable for cases where the CDFs of the two data sets have multiple intersection points, even though they are both monotonic functions. For example, one CDF of a multi-modal distribution may have several ``kinks" or abrupt changes in the slope. Another CDF of a uniform distribution with equal probability for all values is a line with a constant slope. In such cases, our theory is not applicable.

\section{Application to Airless Bodies} \label{sec:airless}

We first apply the theory to examine a simple case---global mean surface temperature on airless bodies. Inhomogeneous illumination patterns occur naturally on a 3D planet, while planetary rotation, radiation, and thermal inertia of the surface and subsurface can modify the energy distribution and inhomogeneity. For an airless body with a uniform surface albedo and emissivity, different rotational states are expected to produce varying mean surface temperatures. According to our theory, a more inhomogeneous body should have a lower mean temperature. To demonstrate this, we first consider three scenarios with different stellation patterns for which analytical temperature solutions can be found. We can then use a surface temperature model to explore the intermediate cases further.
\begin{figure}
  \centering \includegraphics[width=0.45\textwidth]{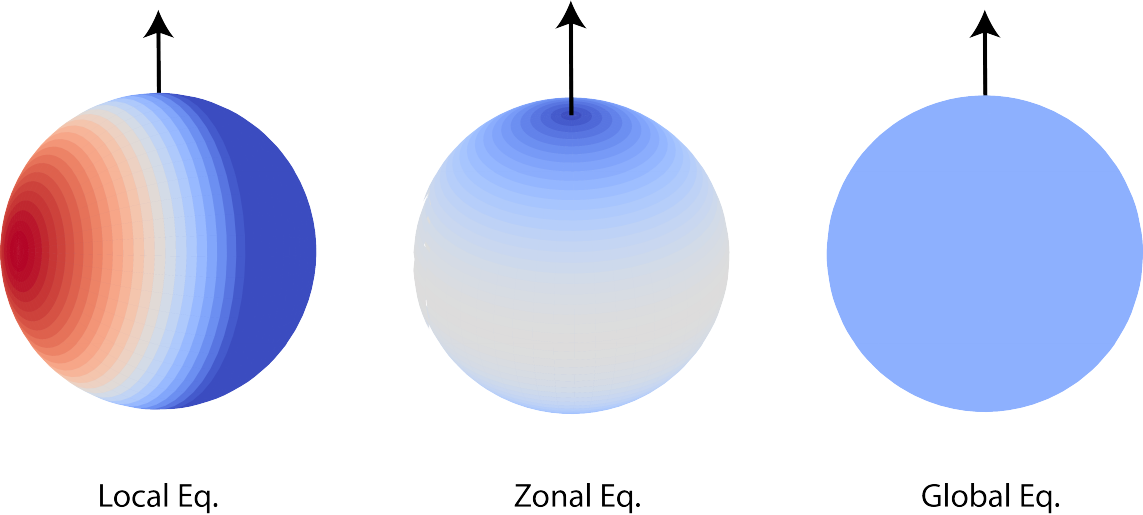} 
  \caption{Illustration of incident stellar flux distribution in the three limiting scenarios, each with analytical solutions: (left) local equilibrium (LE) where the stellar flux is distributed based on the incident angle; (middle) zonal equilibrium (ZE) where the stellar flux is uniformly distributed across longitudes but not latitudes; (right) global equilibrium (GE) where the stellar flux is homogeneously distributed across the entire globe.} \label{cartoon}
\end{figure}

\subsection{Stellar Flux Distributions and CDFs in Three Rotational Configurations}

The three limiting scenarios of stellar flux distribution are illustrated in \Cref{cartoon}. We have assumed a planetary obliquity of zero for simplicity. The local equilibrium (LE) case assumes an instantaneous equilibrium temperature with the incident stellar flux distribution on a slow-rotating or tidally locked planet. The zonal equilibrium (ZE) case assumes a uniformly distributed stellar flux pattern over longitude, which applies to a fast-rotating planet or a surface with large thermal inertia. The global equilibrium (GE) case assumes a globally uniform distribution of stellar flux, which produces the equilibrium temperature via global energy balance. The stellar flux distributions for the three cases as functions of latitude $\phi$ and longitude $\lambda$ are:

\begin{equation}
\begin{split}
    F_{\rm LE}(\lambda,\phi)&=F_0\cos\phi\mathcal{H}(\cos\lambda),\\
    F_{\rm ZE}(\lambda,\phi)&=\frac{1}{2\pi}\int_0^{2\pi}F_0\cos\phi\mathcal{H}(\cos\lambda) d\lambda=\frac{F_0\cos\phi}{\pi}, \\
   F_{\rm GE}(\lambda,\phi)&=\frac{1}{4\pi}\int_{\pi/2}^{\pi/2}\int_0^{2\pi}F_0\cos^2\phi \mathcal{H}(\cos\lambda)d\lambda d\phi = \frac{F_0}{4}, \label{ins}
    \end{split}
\end{equation}
where $F_{\rm LE}$, $F_{\rm ZE}$, and $F_{\rm GE}$ are local incident stellar flux on the sphere for LE, ZE, and GE scenarios, respectively. $F_0$ is the stellar flux at the substellar point. $\mathcal{H}(\cos\lambda)\equiv\max\{\cos\lambda,0\}$ is a Heaviside step function.

Empirically, we can calculate the CDFs of the incident stellar flux distribution using an equal-area grid on a sphere to convert the $\lambda-\phi$ coordinate to the $\lambda-\sin\phi$ coordinate. For the simple three scenarios in \Cref{cartoon}, analytical expressions of CDF can be explicitly derived. In the GE case, the flux is a constant. The CDF is a step function (\Cref{fcdf}):
\begin{equation}
    {\rm CDF}_{\rm GE}(F_{\odot})=
    \begin{cases}
      0, & \text{if}\ F_{\odot}<1/4 \\
      1, & \text{if}\ F_{\odot} \geq 1/4,
    \end{cases}
\end{equation}
where $F_{\odot}$ represents the normalized local incident stellar fluxes by $F_0$ in \Cref{ins}.

For the LE and ZE cases, the probability distribution function (PDF) of the incident stellar flux on a sphere is proportional to the area with the same incident angle $\delta$, such that 
\begin{equation}
    {\rm PDF} (F_{\odot})={\rm PDF} (\delta).
\end{equation}

In the ZE case, the area with the same $\delta$ is proportional to the latitudinal circle length $\cos\phi$. Because $\delta$ equals to the latitude $\phi$ in the ZE case, the PDF of $\delta$ is:
\begin{equation}
    {\rm PDF}_{\rm ZE}(\delta)=\cos\phi=\cos\delta.
\end{equation}
Thus CDF of the stellar flux can be obtained via the integration of the PDF, which gives $1-\sin\delta$. Because  $F_{\odot}=\cos\phi/\pi=\cos\delta/\pi$ (Equation \ref{ins}) in the ZE case, the CDF of the flux $F_{\odot}$ is:
\begin{equation}
    {\rm CDF}_{\rm ZE}(F_{\odot})=1-\left[1-(\pi F_{\odot})^2\right]^{1/2}.
\end{equation}
The CDF shows a curve intersecting the GE line (\Cref{fcdf}).

For the LE case, the stellar flux is zero on the night side. On the dayside, it would be convenient to switch the coordinates to calculate the area of the concentric rings from the substellar point to obtain the PDF of incident angle $\delta$. Because the area of the ring scales as $\sin\delta$, the PDF of $\delta$ over the globe is:
\begin{equation}
    {\rm PDF}_{\rm LE}(\delta)=
    \begin{cases}
      \frac{\sin\delta}{2}, & \text{if}\ \delta \leq \pi/2\\
      0, & \text{if}\ \delta>\pi/2.
    \end{cases}
\end{equation}
We can integrate the above PDF to obtain the CDF of the stellar flux in the LE case. Because $F_{\odot}=\cos\delta$ (Equation \ref{ins}) in the LE case, the CDF of $F_{\odot}$ is:
\begin{equation}
    {\rm CDF}_{\rm LE}(F_{\odot})=\frac{1+F_{\odot}}{2}.
\end{equation}
The CDF shows a slope line after CDF=0.5 (\Cref{fcdf}). 

Each of the CDFs in the three cases only intersects once with one other case (\Cref{fcdf}). From Theorem II in Section \ref{sec:theory}, the stellar flux distribution in the LE case has a larger spatial inhomogeneity than that in the ZE case, which is more inhomogeneous than the GE case.

\subsection{Inequality in Global Mean Surface Temperature}

For an airless body in the above three rotational configurations, the local radiative-equilibrium surface temperature scales with the incident stellar flux as
\begin{equation}
T(F_{\odot})\propto F_{\odot}^{1/4},
\end{equation}
which is a concave function. According to the theory of inhomogeneity effect (Theorem II) in Section \ref{sec:theory}, the global mean surface temperature of the three cases should satisfy the following inequality:
\begin{equation}
\overline{T_{\rm LE}}<\overline{T_{\rm ZE}}<\overline{T_{\rm GE}}, \label{ineqts}
\end{equation}
where $\overline{T_{\rm LE}}$, $\overline{T_{\rm ZE}}$, and $\overline{T_{\rm GE}}$ are global-mean surface temperature for LE, ZE, and GE cases, respectively. 

We can validate \Cref{ineqts} by analytically calculating the mean surface temperature for the three cases. For the GE case, the global-mean temperature is the equilibrium temperature $T_{\rm eq}$:
\begin{equation}
\overline{T_{\rm GE}}=T_{\rm eq}=\left[\frac{(1-A)F_0}{4\epsilon\sigma}\right]^{1/4},
\end{equation}
where $A$ is the bond albedo, $\epsilon$ is the thermal emissivity of the surface, and $\sigma$ is the Stefan-Boltzmann constant.
\begin{figure}
  \centering \includegraphics[width=0.45\textwidth]{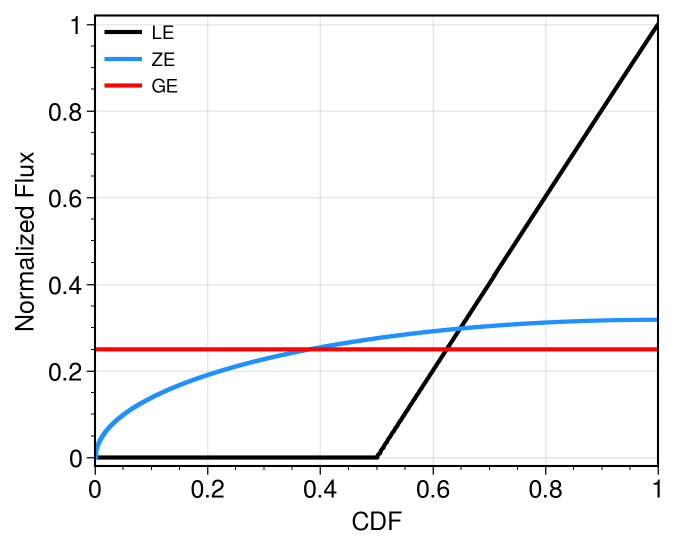} 
  \caption{Cumulative distribution functions (CDFs) of the stellar flux distributions, normalized by the flux at the substellar point, for the LE (black), ZE (blue), and GE (red) cases in \Cref{cartoon}. The LE case is more inhomogeneous than the ZE case. The GE case serves a homogeneous reference.} \label{fcdf}
\end{figure}

For the LE case, \citet{leconte3DClimateModeling2013} obtained the global average by integrating concentric rings from the substellar point. An alternative derivation is provided here:
\begin{equation}
\begin{split}
\overline{T_{\rm LE}}&=\frac{1}{4\pi}\left[\frac{(1-A)F_0}{\epsilon\sigma}\right]^{1/4}\\
&\int_{-\pi/2}^{\pi/2}\cos^{1/4}\phi\cos\phi d\phi\int_0^{2\pi}\mathcal{H}(\cos^{1/4}\lambda) d\lambda.
\end{split}
\end{equation}
Note that
\begin{equation}
\int_{-\pi/2}^{\pi/2}\cos^{x}\phi d\phi=\frac{x-1}{x}\sqrt{\pi}\frac{\Gamma(\frac{x-1}{2})}{\Gamma(\frac{x}{2})}, \label{cosint}
\end{equation}
where $\Gamma(x)=\int_0^\infty t^{x-1} e^{-t}dt$ is the gamma function. We have
\begin{equation}
\overline{T_{\rm LE}}=\frac{2\sqrt{2}}{5} T_{\rm eq}\approx 0.5657~T_{\rm eq}. \label{letem}
\end{equation}
This result is the same as Equation (5) in \citet{leconte3DClimateModeling2013}.

For the ZE case, 
\begin{equation}
\begin{split}
\overline{T_{\rm ZE}}&=\frac{1}{2}\int_{-\pi/2}^{\pi/2}\left[\frac{(1-A)F_0}{\pi\epsilon\sigma}\right]^{1/4}\cos^{1/4}\phi\cos\phi d\phi\\
&=\left[\frac{(1-A)F_0}{16\pi\epsilon\sigma}\right]^{1/4}\int_{-\pi/2}^{\pi/2}\cos^{5/4}\phi d\phi.
\end{split}
\end{equation}
Again, using \Cref{cosint}, we have:
\begin{equation}
\overline{T_{\rm ZE}}=\left(\frac{\pi}{2500}\right)^{1/4}\frac{\Gamma(1/8)}{\Gamma(5/8)} T_{\rm eq}\approx 0.9888~T_{\rm eq}. \label{zetem}
\end{equation}
$\overline{T_{\rm ZE}}$ is slightly smaller than $\overline{T_{\rm GE}}$ by about 1.1\%.
 
The mean surface temperature of the LE case is much less than that in the ZE and GE cases, and the ZE temperature is less than the equilibrium temperature in the GE case (Equation \ref{ineqts}). It confirms our intuition from the theory and the CDFs in \Cref{fcdf}.

\subsection{Mean Surface Temperature Increases with the Rotation Rate and Thermal Inertia}

In paper I, we have shown that the temporally and spatially mean surface temperature of an airless planet is lower than its equilibrium temperature under the global energy balance. Here we can further show that an airless body with a faster rotation or higher thermal inertia should have a higher mean surface temperature, given the same other surface properties. We consider a simple thermophysical model for the surface temperature under periodic stellar forcing. We select one point at the planetary equator and explore the time evolution of the temperature (i.e., the diurnal cycle) and the resultant diurnal-mean temperature. Assuming that the heat flow is transported vertically through the medium via diffusion, we can write the heat equation as:
\begin{equation}
\rho c \frac{\partial T(z,t) }{\partial t} =  \frac{\partial}{\partial z} \left({ \kappa \frac{\partial T(z,t)}{\partial z}}\right),
\end{equation}
where $\rho$ is the density, $c_p$ is the specific heat capacity at constant pressure, $\kappa$ is the thermal conductivity. The boundary condition at the bottom ($z=\infty$) is:
\begin{equation}
\kappa \frac{\partial T(z,t)}{\partial z} \Big|_{\infty} = F_{\rm deep},
\end{equation}
where $F_{\rm deep}$ is the heat flux from the deep subsurface. The surface boundary condition at $z=0$ is:
\begin{equation}
\kappa\frac{\partial T(z,t)}{\partial z} \Big|_0 = \epsilon\sigma T^4(0,t)-(1-A)F(t),
\end{equation}
where $F(t)$ is the time-varying stellar flux. The periodic diurnal forcing can be written as:
\begin{equation}
F(t)=\mathrm{\max}\{F_0\cos(\omega t),0\},
\end{equation}
where $\omega$ is the angular rotation rate, and $F_0$ is the stellar flux at the substellar point. 

For simplicity, we assume $\rho$, $c_p$, and $\kappa$ are invariant with depth and time. For one point at the surface, $\epsilon$ and $A$ are assumed constant with time, i.e., no surface material change via evaporation, sublimation, or condensation. The bottom heat flux $F_{\rm deep}=0$. New variables are introduced to normalize $T$, $z$, $t$, and $F$:
\begin{equation}
\tilde{T}=\frac{T}{T_0},~\tilde{z}=\frac{z}{l_s},~\tilde{t}=\omega t,~\tilde{F}=\frac{F}{F_0},  
\end{equation}
where $T_0$ is the instantaneous equilibrium temperature at the substellar point:
\begin{equation}
T_0=\left[\frac{(1-A)F_0}{\epsilon\sigma}\right]^{1/4}.
\end{equation}
$l_s$ is the diurnal thermal skin depth:
\begin{equation}
l_s=\sqrt{\frac{\kappa}{\rho c_p\omega}}.
\end{equation}
Then the heat equation becomes dimensionless (e.g., \citealt{spencerSystematicBiasesRadiometric1989,selsisEffectRotationTidal2013}):
\begin{equation}
\frac{\partial \tilde{T}(\tilde{z},\tilde{t}) }{\partial \tilde{t}} =  \frac{\partial^2  \tilde{T}(\tilde{z},\tilde{t})}{\partial \tilde{z}^2}.
\end{equation}
The boundary condition at the bottom ($\tilde{z}=\infty$) becomes:
\begin{equation}
\frac{\partial \tilde{T}(\tilde{z},\tilde{t}) }{\partial \tilde{z}} \Big|_{\infty} = 0.
\end{equation}
The surface boundary condition at $\tilde{z}=0$ becomes:
\begin{equation}
\Theta\frac{\partial \tilde{T}(\tilde{z},\tilde{t})}{\partial \tilde{z}} \Big|_0 = \tilde{T}^4(0,\tilde{t})-\mathrm{\max}\left(\cos\tilde{t},0\right),
\end{equation}
where $\Theta$ is the thermal parameter (\citealt{spencerSystematicBiasesRadiometric1989}): 
\begin{equation}
\Theta=\frac{\sqrt{\kappa\rho c\omega}}{\epsilon\sigma T_0^3}, \label{tpar}
\end{equation}
where $\sqrt{\kappa\rho c}$ is the thermal inertia. The system is governed by a single parameter $\Theta$ that considers the thermal inertia of the medium, planetary rotation, and the radiative response of the surface. 

\begin{figure} 
  \centering \includegraphics[width=0.45\textwidth]{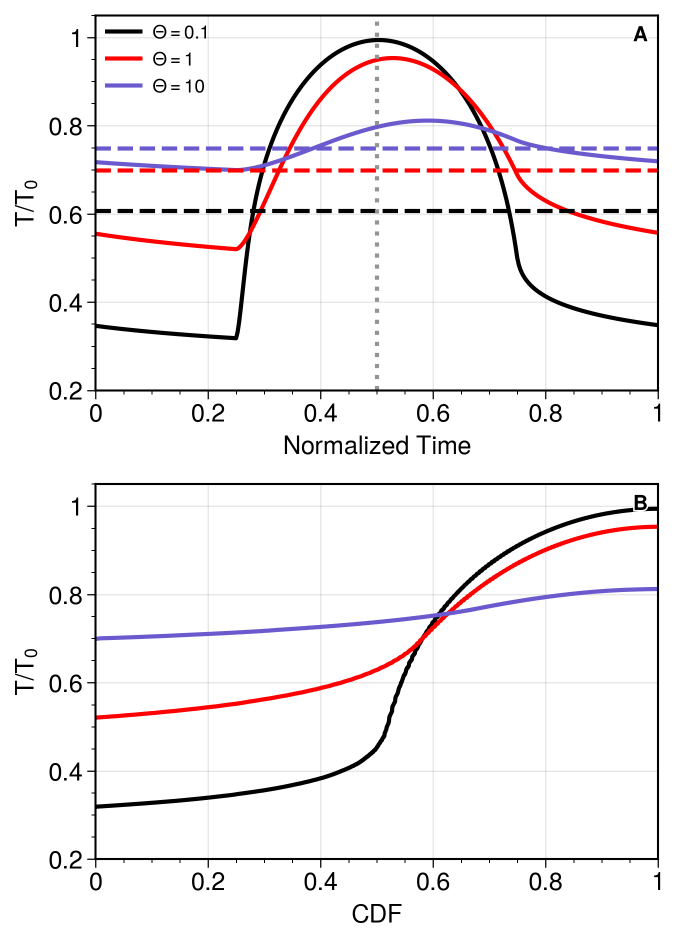} 
  \caption{Surface temperature of an airless body from the thermophysical model. The model is governed by a single parameter $\Theta$. (A): diurnal variation of the normalized temperature ($T/T_0$) for $\Theta=0.1$ (black), 1 (red), and 10 (dark blue). The horizontal axis is normalized time by the rotational period. $t=0.5$ denotes the noon time. The horizontal lines indicate the diurnally averaged temperature. (B): the CDFs of the temperature distributions on the upper panel.} \label{tair}
\end{figure}

We solved the system following the numerical method described in \cite{hayneGlobalRegolithThermophysical2017}. The diurnal evolution of the normalized temperature for typical values of $\Theta$ (0.1, 1, and 10) are plotted in \Cref{tair}A. This plot is similar to Figure 1 in \cite{spencerSystematicBiasesRadiometric1989}. For the case with a larger $\Theta$---indicating a larger thermal inertia or a faster rotation---the temperature variation is smaller, i.e., more homogeneous with time. Moreover, the temperature peak in the diurnal cycle shifts to the afternoon (\Cref{tair}A). A larger $\Theta$ leads to a larger time lag of the temperature peak from the incident stellar flux peak at noon ($t=0.5$). 

To compare different cases, we plotted the CDFs of temperature distributions in \Cref{tair}B. The temperature CDFs from two cases only intersect once, satisfying the criteria described in Section \ref{sec:airless}. The CDF of the case with a larger $\Theta$ appears flatter than the body with a smaller value. According to the theory of inhomogeneity effect, the case with a larger $\Theta$ is more homogeneous. The infrared radiation of the surface blackbody is proportional to $T^4$, which is a convex function. If the diurnal-mean temperatures are the same in these cases, a more homogeneous case must yield a lower total surface cooling flux than a less homogeneous one. However, in a steady state, their total cooling fluxes over a diurnal cycle should be the same. It implies that a more homogeneous case with a larger $\Theta$ should have a higher diurnal-mean temperature. Indeed, the normalized mean surface temperatures (to $T_0$) are 0.6063, 0.6989, and 0.7480 for $\Theta=0.1$, 1, and 10, respectively, confirming the expectation from the theory of inhomogeneity effect. See the horizontal lines in \Cref{tair}A. 

\begin{figure} 
  \centering \includegraphics[width=0.45\textwidth]{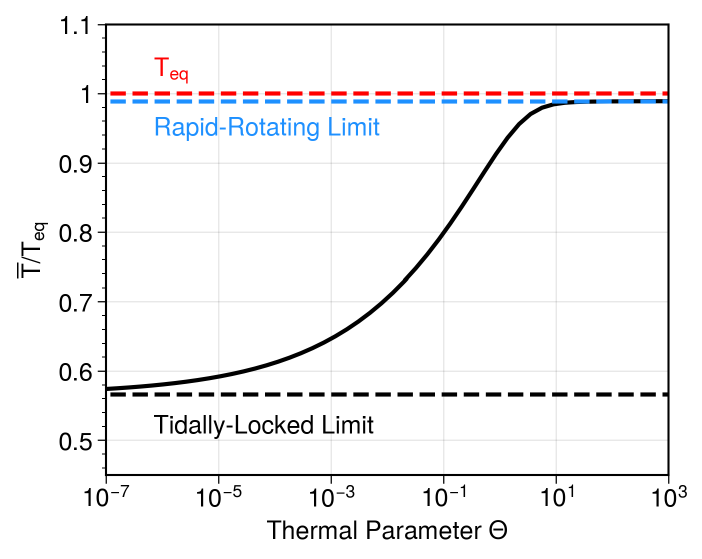} 
  \caption{The ratio (solid) of the diurnal-mean surface temperature to the equilibrium temperature ($\overline{T}/T_{\rm eq}$) as a function of the thermal parameter $\Theta$ assuming a uniform distribution of albedo and emissivity. The dashed line indicates the limiting scenarios in \Cref{cartoon}: black for the tidally locked limit (LE case, $\overline{T}/T_{\rm eq}=2\sqrt{2}/5$), blue for the rapid-rotating limit (ZE case, $\overline{T}/T_{\rm eq}\sim 0.9888$, see Equation \ref{zetem}), and red is the equilibrium temperature (GE case, $\overline{T}/T_{\rm eq}=1$).} \label{tdep}
\end{figure}

The above analysis just focused on a single point on the planet. We can integrate the solution over longitude and latitude to obtain the global-mean and diurnal-mean surface temperatures. We normalized it to the equilibrium temperature in the GE case. The full dependence of the normalized mean temperature on the thermal parameter is illustrated in \Cref{tdep}. As the thermal parameter increases, the mean temperature increases from the tidally-locked limit ($\overline{T}/T_{\rm eq}\sim 0.5657$, Equation \ref{letem}) to the rapid-rotating limit ($\overline{T}/T_{\rm eq}\sim 0.9888$, Equation \ref{zetem}) but it is always smaller than the equilibrium temperature. Because the thermal parameter increases with the thermal inertia and planetary rotation (Equation \ref{tpar}), the mean surface temperature of a non-synchronized rotating, airless body should increase with the surface thermal inertia and rotation rate. The value is bounded by the tidally-locked limit and the rapid-rotating limit.

\section{Dependence of Outgoing Heat Flow of Giant Planets on Rotational States and Heat Redistribution}\label{sec:inso}

To analyze the top-of-the-atmosphere outgoing heat flow of giant planets, we pose an inhomogeneous incident stellar flux pattern on a sphere and solve for the global temperature distribution and the internal heat flux. To ensure everything is analytically trackable, we first consider three limiting scenarios of the rotational states as illustrated in \Cref{cartoon}, where we can find analytical RCE solutions. Then we explore the effect of heat redistribution due to planetary rotation or atmospheric dynamics.

In the LE case, every atmospheric column is in RCE without horizontal heat exchange. This case represents an atmosphere without horizontal dynamics on a slowly rotating planet or in a tidally locked configuration. In the ZE case, the heat redistribution is efficient in the longitudinal direction, and the temperature is uniform along longitudes but not latitudes. It could occur in an atmosphere with a very fast zonal wind or a fast-rotating planet where the incoming stellar flux is averaged out. In the GE case, heat redistribution efficiently homogenizes the temperature across longitude and latitude. The atmosphere can be treated in conventional 1D RCE models.

We will first derive the radiative-equilibrium (RE) temperature distribution in the radiative zone for the three limiting cases and then the RCE solutions in both the radiative and convective zones. 

\subsection{Radiative Equilibrium Solutions}\label{sec:re}

Following the analytical model in Paper I, the RE temperature structure is:
 \begin{equation}
    T_{\rm rad}^4(\tau)=S(\tau)+(D+D^2\tau)F_{\rm int}, \label{trad}
\end{equation}
where $T_{\rm rad}$ is the RE temperature normalized by a reference temperature, which is taken as the equilibrium temperature $T_{\rm eq}$ in this study. $D$ is the diffusivity parameter, $\tau$ is the IR optical depth,  and $F_{\rm int}$ is the outgoing internal heat flux. $S(\tau)$ is the contribution from the stellar heating:
 \begin{equation}
    S(\tau)=\frac{\partial F_{\rm v}}{\partial\tau}-D^2\int_0^\tau F_{\rm v}d\tau-DF_{\rm v}(0).\label{srceq}
\end{equation}
$F_{\rm v}(\tau)$ represents the attenuated stellar flux. All the energy fluxes, including $F_{\rm v}$ and $F_{\rm int}$, are normalized by $4\sigma T_{\rm eq}^4$. The top of the atmosphere incident stellar flux for the three limiting cases is given in \Cref{ins}. Because the flux at the substellar point $(1-A)F_0=4\sigma T_{\rm eq}^4$, the normalized attenuated stellar flux in the LE case is $F_{\rm v}(\tau, \phi, \lambda)=-\mu e^{-\alpha\tau/\mu}$ on the dayside where $\alpha=\kappa_{\rm v}/\kappa$ is the ratio of the visible opacity $\kappa_{\rm v}$ to the infrared (IR) opacity $\kappa$ and $\mu=\cos{\lambda}\cos{\phi}$. $F_{\rm v}=0$ on the night side. The vertical profile of the normalized RE temperature in the LE case $T_{\rm LE}$ is:
\begin{equation}
\begin{split}
    T_{\rm LE}^4(\tau, \phi, \lambda)&=\mu\bigg[D+\frac{D^2\mu}{\alpha}+(\frac{\alpha}{\mu}-\frac{D^2\mu}{\alpha})e^{-\alpha\tau/\mu}\bigg]\\
    &+F_{\rm int,\,LE}(\lambda, \phi)(D+D^2\tau), ~~~~\mathrm{dayside} \\
    T_{\rm LE}^4(\tau, \phi, \lambda)&=F_{\rm int,\,LE}(\lambda, \phi)(D+D^2\tau), ~~~~\mathrm{nightside},
    \label{ret}
\end{split}
\end{equation}
where $F_{\rm int,\,LE}(\lambda, \phi)$ is the normalized internal heat flux measured at the top of the atmosphere in the LE case.

One can show that the RE solution in the ZE case is the zonal average of the LE solution. Taking the zonal-average of the \Cref{ret} by $\int_0^{2\pi}A(\lambda)d\lambda/2\pi$, we obtain the RE temperature profile $T_{\rm ZE}$ in the ZE case:

\begin{equation}
\begin{aligned}
    T_{\rm ZE}^4(\tau,  \phi)&=\frac{\mu_{\phi}}{\pi}\bigg[D+\frac{\pi D^2}{4\alpha}+\frac{\alpha}{\mu_{\phi}}\mathrm{Ki}_1(\frac{\alpha\tau}{\mu_{\phi}})-\frac{D^2\mu_{\phi}}{\alpha}\mathrm{Ki}_3(\frac{\alpha\tau}{\mu_{\phi}})\bigg] \\
    &+F_{\rm int,\,ZE}(\phi)(D+D^2\tau), 
    \label{zeret}
\end{aligned}
\end{equation} 
where $\mu_{\phi}=\cos\phi$ and $F_{\rm int,\,ZE}(\phi)$ is the normalized internal heat flux at the top of the atmosphere in the ZE case. $\mathrm{Ki}_n(x)$ is the Bickley-Naylor function $\mathrm{Ki}_n(x)=\frac{x^n}{(n-1)!}\int_1^\infty (t-1)^{n-1}K_0(xt)dt$ where $K_0$ is the Bessel function of the second kind (\citealt{bickleyXXVShortTable1935,abramowitzHandbookMathematicalFunctions1964}). 

Likewise, the solutions in the GE case are the global average of LE solutions. Because the LE solution (Equation \ref{ret}) only depends on $\mu=\cos\lambda\cos\phi$, we can perform the global average by integrating concentric rings from the substellar point by $\int_{-1}^{1}A(\mu)d\mu/2$. The RE temperature profile in the GE case $T_{\rm GE}$ is given by:

\begin{equation}
\begin{split}
    T_{\rm GE}^4(\tau)&= \frac{1}{2}\bigg[\frac{D}{2}+\frac{D^2}{3\alpha}+\alpha E_2(\alpha\tau)-\frac{D^2}{\alpha}E_4(\alpha\tau)\bigg]\\
    &+F_{\rm int,\,GE}(D+D^2\tau),
    \label{geret}
\end{split}
\end{equation} 
where $E_n(x)$ is the exponential integral and $F_{\rm int,\,GE}$ is the normalized internal heat flux at the top of the atmosphere in the GE case. If we adopt $D=\sqrt{3}$, the global-mean RE temperature profile is consistent with the global-mean solutions under the Eddington approximation in previous studies. For example, we can reproduce Equation (49) in \cite{guillotRadiativeEquilibriumIrradiated2010} and Equation (A11) in \cite{zhangRadiativeForcingStratosphere2013}  if we chose the second Eddington coefficient $f_{\rm Hth}=1/\sqrt{3}$ in their solutions. 

We define the global-mean internal heat fluxes in the three limiting cases respectively:
\begin{equation}
\begin{split}
    \overline{F_{\rm int,\,GE}}&=F_{\rm int,\,GE}, \\
    \overline{F_{\rm int,\,ZE}}&=\frac{1}{2}\int_{-\pi/2}^{\pi/2}F_{\rm int,\,ZE}(\phi)\cos\phi d\phi, \\
    \overline{F_{\rm int,\,LE}}&=\frac{1}{4\pi}\int_{-\pi/2}^{\pi/2}\int_0^{2\pi}F_{\rm int,\,LE}(\lambda, \phi)\cos\phi d\lambda d\phi.
    \end{split}
\end{equation}
Because the global-mean RE temperature profiles of the LE and ZE cases are the same as the GE solution, the globally averaged internal heat fluxes are also the same in all three cases under \textit{radiative equilibrium}:
\begin{equation}
\overline{F_{\rm int,\,LE}}=\overline{F_{\rm int,\,ZE}}=\overline{F_{\rm int,\,GE}}. \label{feqre}
\end{equation}

We can generalize this argument from three limiting scenarios to any conservative horizontal heat redistribution. Take the horizontal average of the radiative transfer equation (Equation 6 in Paper I):
\begin{equation}
\frac{d^2\overline{F}}{d\tau^2}=D^2\overline{F}-\frac{d\overline{T^4}}{d\tau}, 
\end{equation}
where $F$ is the net infrared flux. In a steady state, the heating and cooling at every atmospheric level balance:
\begin{equation}
\frac{\partial (\overline{F}+\overline{F_{\rm v}})}{\partial \tau}+\overline{q}=0,
\end{equation}
where $q$ represents the heating/cooling rate due to horizontal heat redistribution such as planetary rotation, heat advection, and wave mixing. As long as the redistribution process is conserved on the horizontal plane, i.e., $\overline{q} = 0$, the global-mean radiative solution of the temperature ($\overline{T^4}$) remains the same, and so is the global-mean internal heat flux ($F_{\rm int}$). This explains \Cref{feqre} but can be applied to any horizontally conservative heat transport process. This argument has been addressed in the RE calculations in \cite{guillotRadiativeEquilibriumIrradiated2010}. However, as demonstrated in the following section, the equalities presented in \Cref{feqre} fail to hold in RCE cases with underlying convective regions.

\begin{figure*}
  \centering \includegraphics[width=0.99\textwidth]{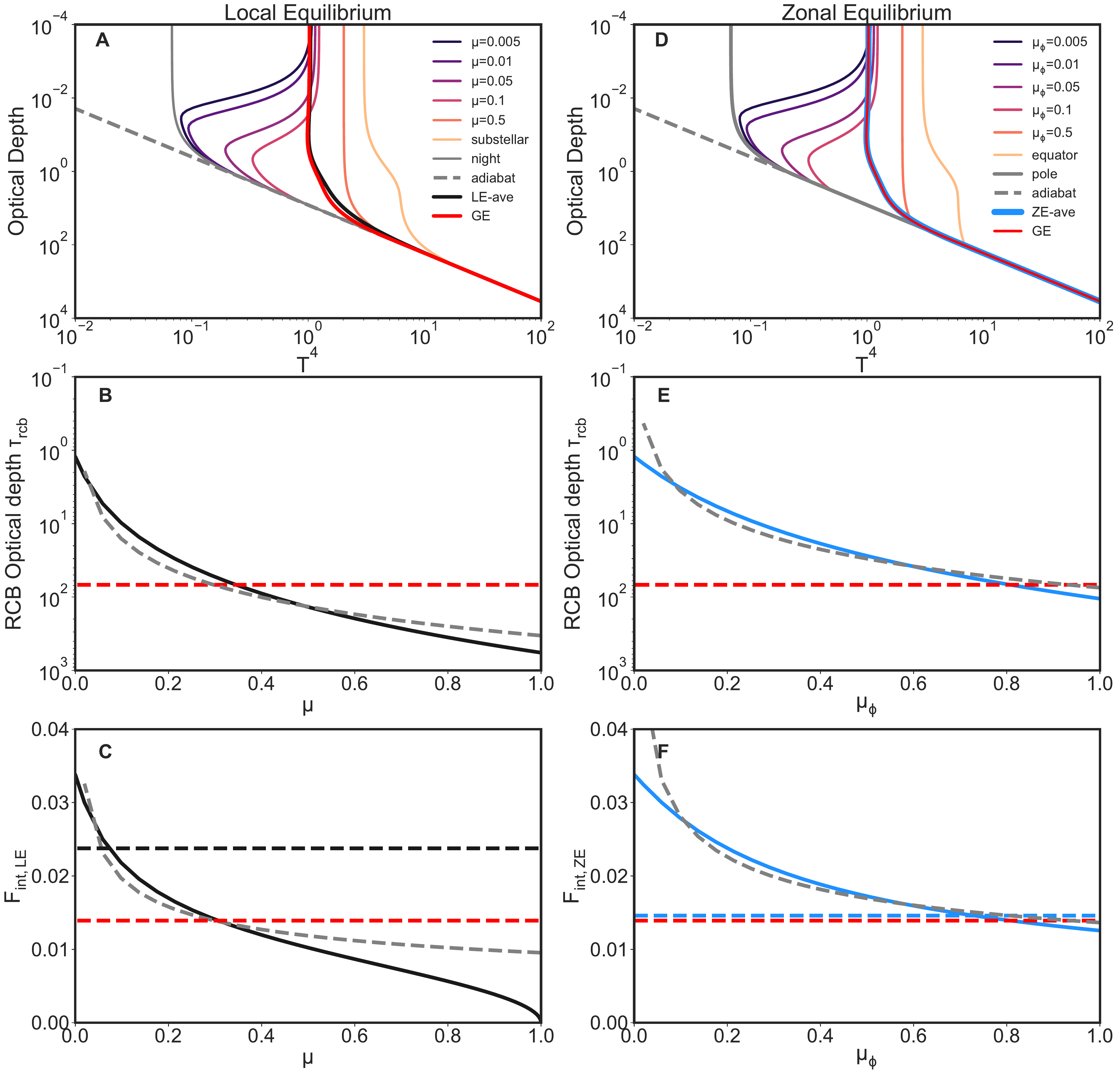} 
  \caption{RCE solutions of the LE (left panels), ZE (right panels), and GE (in both panels) cases with $\beta\sim 0.76$, $D=2$, $K=0.2$, and $\alpha=1$. (A): Temperature profiles from the LE cases at selected incident angles. The globally averaged profile in the LE case (black) is compared with that in the GE case (red). (B): The RCB optical depth as a function of the cosine of the stellar incident angle ($\mu$) from the LE case (black), GE case (red dashed), and an analytical solution (gray dashed) using \Cref{anafint}. The nightside corresponds to $\mu=0$. (C): The normalized internal heat flux ($F_{\rm int,\,LE}$) as a function of $\mu$. The global-mean heat flux from the LE case (black dashed) is compared with that in the GE case (red dashed), and an analytical solution (gray dashed) using \Cref{anafint}. (D): Temperature profiles from the ZE cases at selected incident angles (indicating different latitudes). The globally averaged profile in the ZE case (blue) is almost the same as in the GE case (red). (E): The RCB optical depth as a function of $\mu$ from the ZE case (blue), GE case (red dashed), and an analytical solution (gray dashed) using \Cref{anafint}. (F): The normalized internal heat flux ($F_{\rm int,\,ZE}$) as a function of $\mu$. The global-mean heat flux from the ZE case (blue dashed) is slightly larger than that in the GE case (red dashed). The analytical solution (gray dashed) from \Cref{anafint} is also shown.} \label{insofig}
\end{figure*}

\subsection{Radiative-Convective Equilibrium Solutions}\label{secrce}

The temperature in the convective zone, $T_{\rm conv}$, follows the adiabatic relation below the radiative-convective boundary (RCB). 
\begin{equation}
T_{\rm conv}^4(\tau)=K\tau^\beta,
\label{tconv}
\end{equation}
where $\beta$ is the adiabatic slope parameter. $\beta$ is approximately 0.76 in a dry, hydrogen-dominated atmosphere (see the discussion in Paper I). We assumed $\beta$ is a constant in this study, but it can vary locally depending on the adiabatic index and opacity distribution in the real atmosphere. The interior temperature proxy parameter $K=T_{\rm s}^4\tau_{\rm s}^{-\beta}$, measures how hot the interior is. $T_{\rm s}$ is the temperature in the deep convective zone at a reference optical depth $\tau_{\rm s}$. The RCB is determined using the temperature and flux continuity constraints outlined in Paper I.

The RCE solutions of the vertical temperature profiles for LE, ZE, and GE cases are presented in Figures \ref{insofig}A and \ref{insofig}D. We used $\beta\sim 0.76$, $D=2, ~K=0.2$, and $\alpha=1$ in the $F_{\rm v}$. The LE temperature decreases from the substellar point ($\mu=1$) to the terminator/nightside ($\mu=0$). Some dayside temperature profiles with large $\mu$ show a vertical inversion in the radiative zone due to strong stellar heating. The dependence of the temperature profile on the stellar incident angle in the radiative zone is similar to the results from pure RE calculations in \cite{guillotRadiativeEquilibriumIrradiated2010} (as shown in their Figure 3). 

For LE and ZE cases, we also calculated the area-weighted globally-averaged temperature profiles via ${(\overline{T^4})}^{1/4}$ to conserve the energy. The averaged temperature in the LE case is hotter than the GE case above the RCB, indicating a larger outgoing flux at the top of the atmosphere (\Cref{insofig}A), which is further evidenced by a larger internal heat flux in the LE case (\Cref{insofig}C). On the other hand, the globally averaged ZE temperature is only slightly hotter than the GE case. The difference is less than a few percent and thus is not visible in \Cref{insofig}D. Accordingly, the internal heat flux in the ZE case is also very close to that in the GE case (\Cref{insofig}F). 

The dependence of the RCB and local interior heat flux ($F_{\rm int,\,LE}$) on the incident angle is shown in \Cref{insofig}B and \Cref{insofig}C, respectively. From the night side (or the poles) to the substellar point, the RCB moves deeper into the atmosphere, and the emitted local interior heat flux decreases as $\mu$ increases. The $\mu$ dependence of the ZE solution is similar to the LE case. From the equator ($\mu_{\phi}=1$) to the poles ($\mu_{\phi}=0$), the ZE temperature decreases, the RCB level moves higher, and the interior heat flux reduces. In the GE scenario, the temperature profile, location of the RCB, and internal heat flux are all confined within the extremes delineated by the direct incident beam case ($\mu$ or $\mu_{\phi}=1$) and the nightside/polar case ($\mu$ or $\mu_{\phi}=0$). These limits are shown by the red lines in \Cref{insofig}.

The internal heat flux tends to escape more at the poles and on the night side because the vertical gradient of the temperature profile at these places is steeper, and the RCB is higher, allowing more flux to be emitted (\citealt{guillotEvolution51Pegasus2002}). \citet{pierrehumbertThermostatsRadiatorFins1995} coined the term ``radiator fins" to depict dry regions in Earth's tropics and sub-tropics. These areas readily lose longwave infrared flux to space due to low water opacity. This concept has also been applied to explain the cooling effect on the drier night side of tidally locked terrestrial planets (e.g., \citealt{yangLOWORDERMODELWATER2014}). In our model of dry giant planets, we use this analogy to underscore the significance of the poles and the night side of tidally locked planets. These areas, due to their steeper temperature gradients, permit a greater amount of \textit{internal heat flux} to escape into space. In our next paper (\citealt{zhangInhomogeneityEffectIII2023}), we will extend the ``radiator fins" analogy to include opacity inhomogeneity, thus illustrating the impact of temperature-dependent opacity on the interior cooling of tidally locked gas giants.

\begin{figure*}
\centering \includegraphics[width=0.95\textwidth]{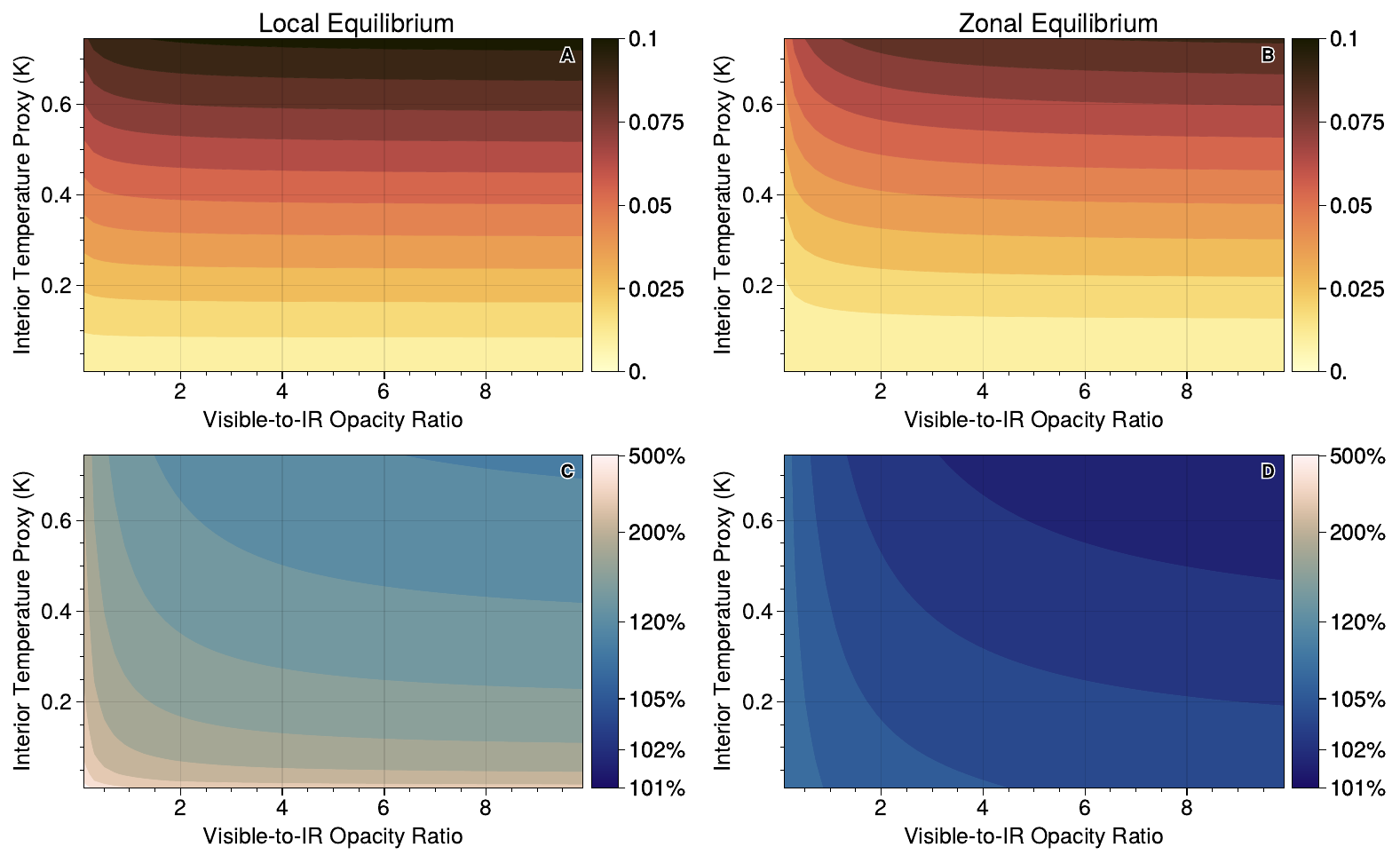}
\caption{Upper row: the global average internal heat flux as functions of the visible-to-IR opacity ratio $\alpha$ and the interior temperature proxy $K$ in the LE (panel A) and ZE cases (panel B). Lower row: the ratio of internal heat in the LE (panel C) and ZE (panel D) cases to the GE case. Note that the contour and color bar grids in Panels C and D are non-uniform to enhance detail.} \label{akcontour}
\end{figure*}

Our RCE model does not provide explicit solutions for RCB optical depth and internal heat flux (refer to Equations 37 and 38 in Paper I). Using temperature continuity and the Schwarzschild criterion, we derived approximate expressions for $\tau_{\rm rcb}$ and $F_{\rm int}$ for irradiated gas giants in Appendix \ref{app:anarcb} (see Equation \ref{anarcb}):
\begin{equation}
\begin{split}
\tau_{\rm rcb}&\sim\left[\frac{2DF_{\odot}}{(1-\beta)K}\right]^{1/\beta},\\
F_{\rm int}&\sim\frac{\beta K^{1/\beta}}{D^2}\left(\frac{1-\beta}{2DF_{\odot}}\right)^{(1-\beta)/\beta}.
\end{split}
\label{anafint}
\end{equation}
where $F_{\odot}$ represents the local incident stellar flux. In normalized terms, $F_{\odot}=\mu$ for the LE case (dayside), $F_{\odot}=\mu_{\phi}/\pi$ for the ZE case, and $F_{\odot}=1/4$ for the GE case.

The analytical expressions reveal the quantitative dependence of the RCB and internal heat flux on the incident stellar flux. As the stellar flux increases, the RCB should shift deeper and inhibits the outgoing internal heat flux. This is especially true when the incident beam is more direct (i.e., larger $\mu$ or $\mu_{\phi}$). With $D=2$ and $\beta\sim 0.76$, our simple estimate (represented by the grey dashed lines) aligns well with the trends of RCB location and internal heat flux in relation to the incident angle for both LE (\Cref{insofig}B and C) and ZE cases (\Cref{insofig}E and F). For the GE case, our estimated $\tau_{\rm rcb}\sim 37.84$ is within a factor of two of the semi-analytical RCE model result of $68.17$ (shown as red dashed lines in \Cref{insofig}B and E). Our estimated internal heat flux $F_{\rm int}\sim 0.01471$ is close to the RCE model result of 0.01392 (illustrated by red dashed lines in \Cref{insofig}C and F).

To apply the theory of the inhomogeneity effect, we should examine the curvature of the outgoing internal flux at the top of the atmosphere in relation to the incoming stellar flux. Strictly speaking, the exact value of the outgoing internal heat flux depends not only on the top-of-the-atmosphere stellar flux but also on how the flux is attenuated through the atmosphere. However, as shown by a good match between our analytical solution of $F_{\rm int}$ (Equation \ref{anafint}) and the semi-analytical RCE model results (\Cref{insofig}), $F_{\rm int}$ should be primarily determined by the magnitude of the incident stellar flux $F_{\odot}$ rather than the shape of the flux attenuated profile. 

Given an internal entropy (i.e., fixing $K$), the analytical solution of $F_{\rm int}$ (Equation \ref{anafint}) implies:
\begin{equation}
F_{\rm int}\propto F_{\odot}^{(\beta-1)/\beta}.
\end{equation}
Because $\beta < 1$, the dependence of $F_{\rm int}$ on $F_{\odot}$ is a convex function (also see Figure 5A in Paper I). As $F_{\odot}$ is proportional to $\mu$ in the LE case and $\mu_{\phi}$ in the ZE case, respectively, the convexity from our more accurate RCE model is also pronounced in \Cref{insofig}C and \Cref{insofig}F. In the LE case, the function is slightly concave when $\mu$ is very close to unity (i.e., at the substellar point, \Cref{insofig}C), but since the area is very small, we neglect its contribution to the total convexity. 

The CDFs of the stellar flux distribution in the three limited cases are the same as in the airless cases shown in \Cref{fcdf}. According to the theory of inhomogeneity effect (Theorem II) in Section \ref{sec:theory}, the global-mean outgoing internal flux of giant planets in the LE case should be the largest, that in the GE case be the lowest, and that in the ZE case be in the middle: 
\begin{equation}
\overline{F_{\rm int,\,LE}}>\overline{F_{\rm int,\,ZE}}>\overline{F_{\rm int,\,GE}}. \label{ineqrce}
\end{equation}

Our calculated global-mean internal fluxes in the three limiting RCE cases confirmed the above inequality: $\overline{F_{\rm int,\,LE}}\sim 0.02377$ (black dashed line in \ref{insofig}C), $\overline{F_{\rm int,\,ZE}}\sim 0.01460$ (blue dashed line in \ref{insofig}F), and $\overline{F_{\rm int,\,GE}}\sim 0.01392$ (red dashed lines in \ref{insofig}C and \ref{insofig}F). Different from the equality in the RE cases (Equation \ref{feqre}), the global-mean internal heat fluxes in the RCE cases show that inhomogeneity of the incident stellar flux enhances the outgoing internal heat flux.

With our adopted parameters, the LE flux is about 70\% larger than the GE flux, while the ZE flux is slightly higher than the GE flux. Averaging the heat flux between the dayside and the nightside results in a large decrease in the internal heat flux from the LE to ZE cases. But in the ZE case, due to a smaller polar area than the equatorial region, the outgoing heat flux is significantly weighted towards the equator ($\mu_{\phi}=1$) where the dependence of the internal heat flux on the incident stellar flux is almost linear (\Cref{insofig}F). The latitudinal average does not result in a significant change in the global-mean internal heat flux from the ZE to the GE case. Although the nightside and the polar region behave as the ``radiator fins" of the planet, the nightside plays a larger role in cooling the planet than the polar region due to its larger area.

We validated the inequality in \Cref{ineqrce} across a wide range of parameters by varying $\alpha$ from 0.1 to 10 and $K$ from 0.01 to 0.75. A larger visible-to-IR opacity ratio $\alpha$ means that the stellar flux is absorbed in a higher altitude, resulting in less inhibition of internal heat flux. A larger $K$ implies a hotter interior and internal heat flux. The normalized global-mean internal heat flux, $\overline{F_{\rm int,\,LE}}$, ranges from 0.001 to 0.1 (\Cref{akcontour}A). The global-mean internal flux in the LE case is strongly influenced by the interior temperature (represented by $K$) but only weakly decreases with increasing $\alpha$. The internal flux in the ZE case shows a similar pattern, with a stronger dependence on $\alpha$ and weaker dependence on $K$ than the LE case. The ratio of the global-mean flux in the LE case to the GE case ranges from 110\% to 600\%, and that in the ZE case to the GE case ranges from 101\% to 110\% in this range of parameters (\Cref{akcontour}B). Although this survey confirms that the inhomogeneity effect holds in large parameter space, the day-night stellar flux pattern significantly enhances the interior cooling of giant planets more than the equator-to-pole irradiation pattern.

The inhomogeneous incident stellar flux has a greater impact on the internal heat flux when the stellar flux can penetrate deeper into the atmosphere. As a result, the ratio of internal heat flux in the LE to the GE case decreases as $\alpha$ and $K$ increase (\Cref{akcontour}B). If $\alpha$ is 0.1 and $K$ is 0.01, the stellar flux can reach very deep into the atmosphere, resulting in the internal heat flux in the LE case being approximately six times that of the GE case. Even when both $\alpha$ and $K$ are large, the inhomogeneity effect still makes a notable difference between the LE and GE fluxes.

According to 1D internal structure models and current mass-radius data of hot Jupiters, the intrinsic temperature ($T_{\rm int}$)---the brightness temperature producing the required internal heat flux---can vary from 100 to 700 K, with equilibrium temperatures ranging from 800 to 2500 K \citep{thorngrenIntrinsicTemperatureRadiative2019}. The corresponding normalized internal heat flux---equivalent to the required heating efficiency from an unknown heating mechanism to explain the large radii of hot Jupiters--- ranges from 0.0001 to 0.025 (\citealt{thorngrenBayesianAnalysisHotJupiter2018,sarkisEvidenceThreeMechanisms2021}). As shown in \Cref{akcontour}A, these data correspond to $K<0.2$. Within this range, the flux in the LE case is at least 50\% larger than the GE flux for highly inflated hot Jupiters ($K\sim 0.2$, \Cref{akcontour}B), and the ratio of LE to GE flux increases as $K$ decreases. For less inflated hot/warm Jupiters, both $K$ and internal flux are small. In this scenario, the 1D GE model might underestimate the internal heat flux several times compared to the LE case. The effect of inhomogeneity remains important until a very high intrinsic temperature. For example, if the intrinsic temperature is 1000 K, the normalized internal heat flux is about 0.1 for a planet with $T_{\rm eq}\sim 1800$ K. The difference between the LE and GE cases is only about 10\%. This could happen in the early contraction phase of hot planets. Note that the above discussion is based on a crude grey atmospheric treatment, and the real planetary atmosphere is not in an LE state due to dynamics. The more realistic effect of inhomogeneity on planetary evolution needs to be studied in future work.

In sum, the inhomogeneous distribution of incident stellar flux on the sphere and a strong day-night contrast significantly increases the outgoing planetary heat flow. The global-mean 1D calculation (GE) underestimates the heat flow for a planet with significantly day-night inhomogeneity with inefficient heat redistribution (LE). But for the fast-rotating planets or efficient zonal heat redistribution (ZE), the global-mean 1D model might be a good assumption. In the following discussion, we discuss the effect of rotation and dynamics with intermediate heat redistributed efficiency. 

\subsection{Effect of Horizontal Heat Redistribution}\label{advsec}

Unlike the LE state with instantaneous equilibrium and the ZE or GE state assuming complete homogenization, the intermediate state with moderate heat redistribution efficiency is more realistic. In the atmosphere above the RCB, horizontal homogenization can be approximated by redistributing the incoming stellar energy via non-synchronous rotation, horizontal wind advection, and eddy heat mixing. As shown in Section \ref{sec:re}, if there is no convective layer and the vertical heat transport is only controlled by radiation, any horizontally conservative heat transport process cannot alter the global-mean value of the internal heat flux. However, this is not true when convection is present. Even a horizontally conservative heat redistribution process can produce different outgoing heat fluxes. An example has been shown in \Cref{ineqrce} by directly averaging the incident stellar flux over space, which is a conservative redistribution process. 

To demonstrate the generic effect of heat redistribution, we introduced an efficiency factor to characterize the effect of heat transport or planetary rotation. Previous studies have used different definitions of heat redistribution efficiency in literature (e.g., see discussion in the appendix in \citealt{spiegelATMOSPHERESPECTRALMODELS2010}). In this study, we define a zonal heat redistribution efficiency $\eta$ ($0\le \eta \le 1$) to linearly combine the incident stellar flux contribution between the LE and ZE cases by modifying the attenuated stellar flux $F_{\rm v\eta}$:
\begin{equation}
F_{\rm v\eta}(\tau, \phi, \lambda) = (1-\eta)F_{\rm v, LE}(\tau, \phi, \lambda)+\eta F_{\rm v, ZE}(\tau,\phi),
\end{equation}
where $F_{\rm v, LE}$ represents the distribution of stellar flux in the LE case and $F_{\rm v, ZE}$ is in the ZE case. As $F_{\rm v}$ changes, its vertical integral, vertical derivative, and $S(\tau)$ (Equation \ref{srceq}) also change accordingly. The horizontal heat redistribution is conservative in this design because $\overline{F_{\rm v\eta}}=\overline{F_{\rm v, LE}}=\overline{F_{\rm v, ZE}}$.
\begin{figure}
  \centering \includegraphics[width=0.45\textwidth]{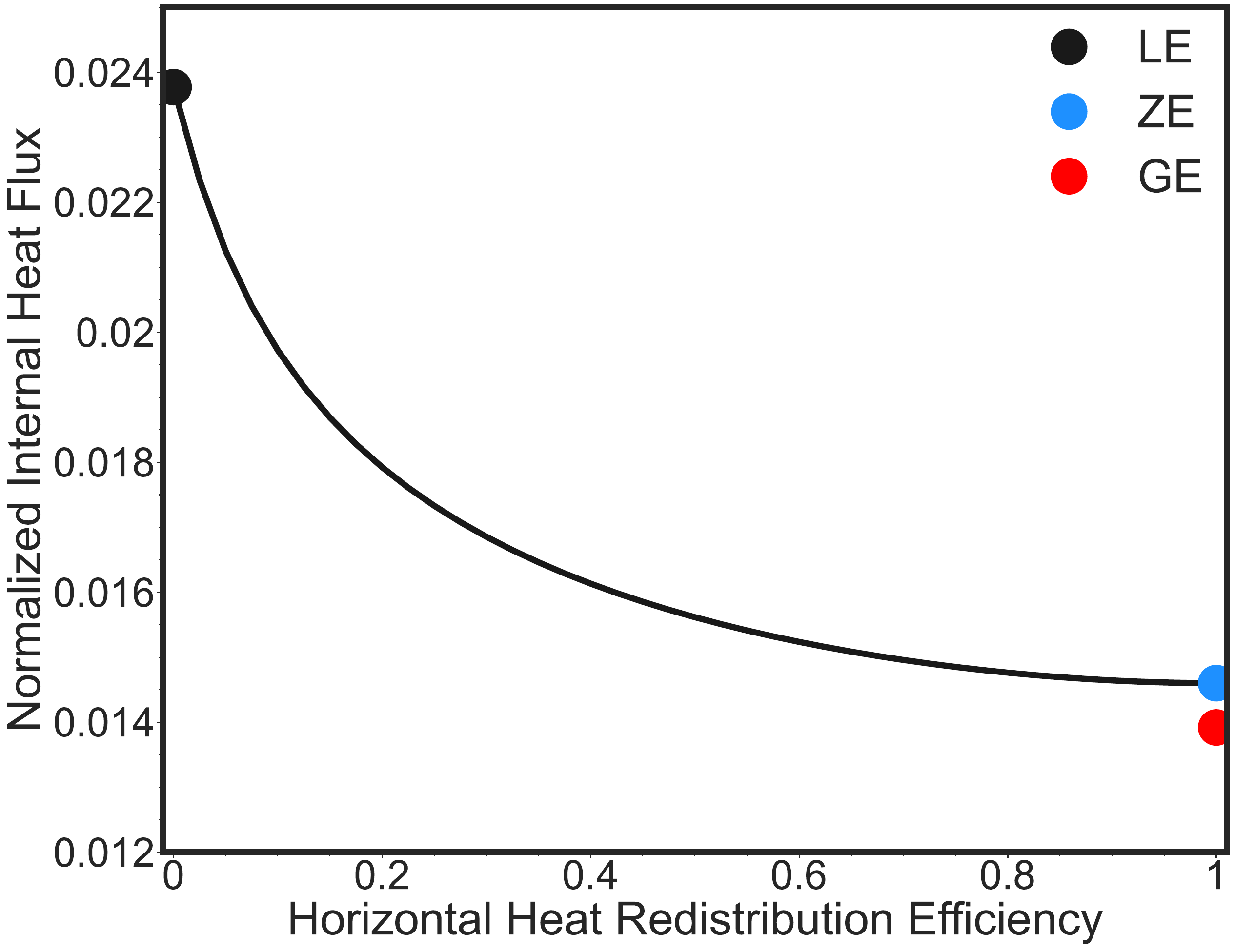} 
  \caption{Normalized global-mean internal heat flux as a function of horizontal heat redistribution efficiency $\eta$. The basic parameters are the same as the LE case in \Cref{insofig}.  When the efficiency is zero (no redistribution), the internal heat flux is the same as the LE case (black dot). As the efficiency approaches unity (horizontal homogeneity), the internal heat flux converges to the ZE case (blue dot). For reference, the flux in the GE case (red dot) is slightly lower than in the ZE case.} \label{advf} 
\end{figure}

We then use the RCE model to solve for the temperature profiles and the outgoing internal heat fluxes at the top of the atmosphere. By varying $\eta$ from 0 to 1, we can systematically explore the dependence of the interior cooling flux on heat redistribution.

\begin{figure}
  \centering \includegraphics[width=0.45\textwidth]{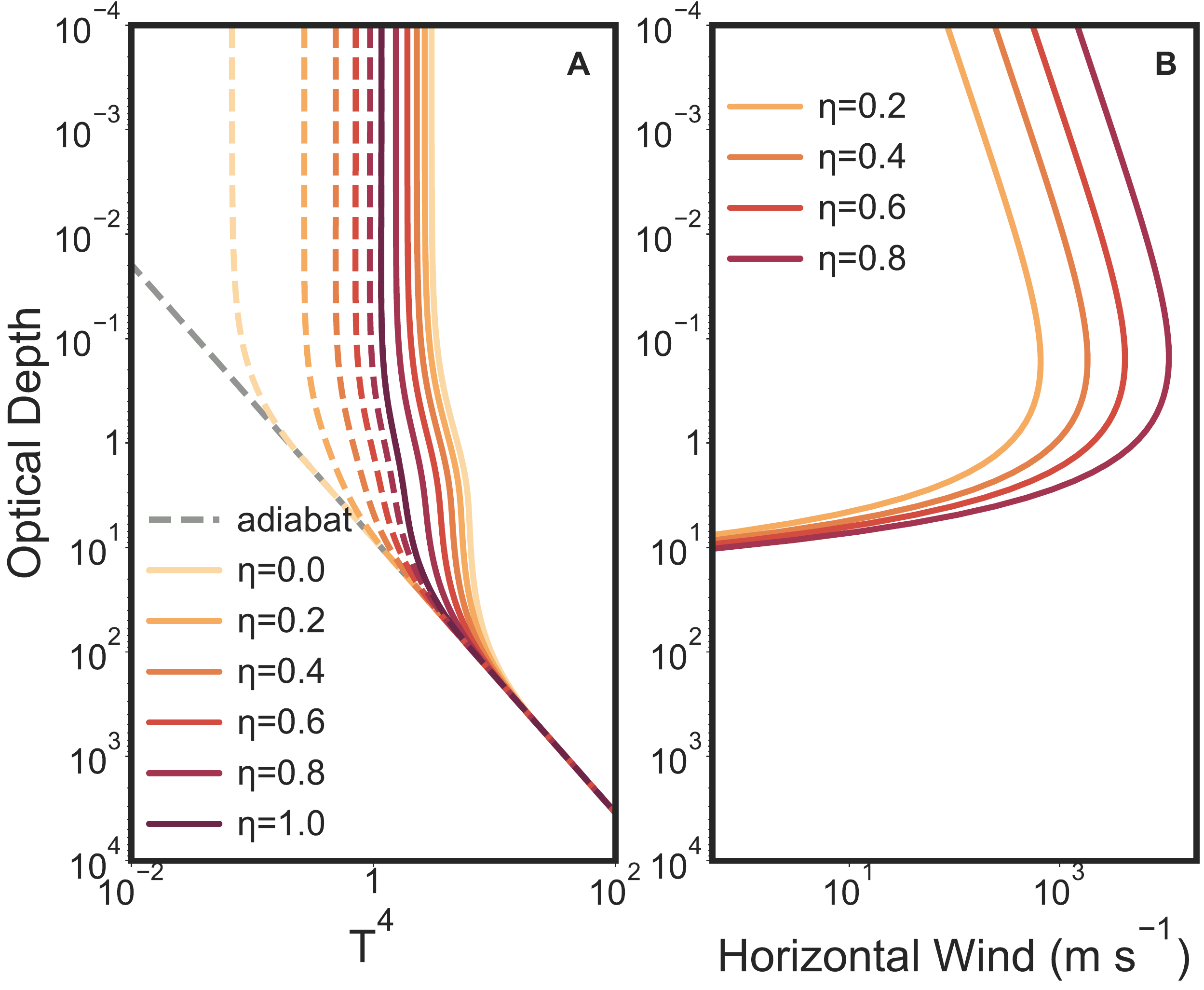} 
  \caption{Normalized temperature profiles (A) and the derived horizontal wind profile (B) as a function of horizontal heat redistribution efficiency ($\eta$). The wind profiles are not shown for zero and unity efficiencies, as they are ill-defined.}\label{advt} 
\end{figure}

As heat redistribution efficiency increases, the horizontal inhomogeneity between day and night decreases. The internal heat flux also decreases with $\eta$ (\Cref{advf}). It decreases faster in the low-$\eta$ regime and slower in the high-$\eta$ regime. The temperature profiles at the substellar and anti-stellar points are shown in \Cref{advt}. If $\eta=0$, it is similar to the local equilibrium case with the largest temperature difference between the substellar and anti-stellar points (see lines in \Cref{insofig}A). If $\eta=1$, both the dayside and nightside have the same temperature profiles, and the solutions converge to the ZE case.

If the heat redistribution is primarily caused by horizontal atmospheric processes such as jet advection and wave mixing on tidally locked exoplanets, we can estimate the distribution of the effective wind speed. In a steady state, the heat transport equation requires that the horizontal heat flux balances the local net radiative heating in the radiative zone:
\begin{equation}
\vec{\bf{u}}\cdot\nabla_pT = \frac{g}{c_p}\frac{\partial F_{\rm rad}(p, \phi, \lambda)}{\partial p},\label{adveq}
\end{equation}
where $p$ is pressure, $\nabla_p$ is the gradient operator on an isobar, and $c_p$ is heat capacity at constant pressure. $\vec{\bf{u}}$ is the effective horizontal wind velocity that could represent the jet speed in advection or eddy wind speed in mixing. $F_{\rm rad}$ is the net vertical radiative flux. In our model, the net infrared flux is $-F_{\rm v\eta}+F_{\rm int}$, and the visible flux is the instantaneous radiation $F_{\rm v, LE}$. Thus $F_{\rm rad}=F_{\rm v, LE}-F_{\rm v\eta}+F_{\rm int}=\eta(F_{\rm v, LE}-F_{\rm v, ZE})+F_{\rm int}$. $F_{\rm rad}$ is varying with longitude. As an order of magnitude calculation, we evaluated the magnitude of $F_{\rm rad}$ at the anti-stellar point: $F_{\rm rad}\approx-\eta F_{\rm v, ZE}+F_{\rm int}$. We also approximated $\vec{\bf{u}}\cdot\nabla_pT\approx -U \Delta T$ where $U$ is the wind speed and $\Delta T$ is the temperature difference between the substellar and anti-stellar points. The negative sign indicates that the heat flux is transported toward the cooler anti-stellar point.

We assume $\tau=\kappa p^n$ to convert the vertical coordinate from $p$ to $\tau$. \Cref{adveq} can be transformed into the wind equation: 
\begin{equation}
U(\eta, \tau)= \eta\frac{gn \kappa^{1/n}\tau^{(n-1)/n} R_p}{c_p \Delta T(\eta, \tau)}\frac{\partial \left[F_{\rm v, ZE}(\tau,\phi)+F_{\rm int}\right]}{\partial\tau},
\end{equation}
where $R_p$ is the planetary radius. The vertical gradient of the constant $F_{\rm int}$ is zero. For a solar-metallicity and hydrogen-dominated atmosphere at about 2000 K \citep{freedmanGaseousMeanOpacities2014}, $n\approx 1.5$ and $\tau\approx 10^{-5}p^{1.5}/g$ where $p$ is pressure in $\mathrm{Pa}$ and $g$ is the gravity in $\mathrm{m~s^{-2}}$. For a typical hot Jupiter, we used $g=10~ \mathrm{m~s^{-2}}$, $R_p=10^8~\mathrm{m}$, and $c_p\sim10^4 ~\mathrm{J~Kg^{-1}~K^{-1}}$. 

The magnitude of the derived zonal wind (\Cref{advt}) is consistent with previous GCM studies for hot Jupiters (e.g., \citealt{showmanAtmosphericCirculationHot2009}). The maximum wind velocity occurs around the optical depth unity level. The wind velocity is zero for the $\eta=0$ (LE) case. The peak wind velocity spans from a few hundred $\mathrm{m~s^{-1}}$ in the low-efficiency case to a few thousand $\mathrm{m~s^{-1}}$ in the high-efficiency case. The wind velocity rapidly decreased in the high optical depth region. However, our analysis did not provide information on winds below the RCB, as a single adiabat is used for the convective region. Assuming the heat redistribution efficiency $\eta$, we avoided the direct calculation of $\nabla_{p}T$. This method resulted in a wind distribution $\vec{\bf{u}}$ that may not be consistent with the temperature distribution. A self-consistent solution for the wind and temperature would require solving the atmospheric dynamics, which is beyond the current analytical framework.

\section{Dependence of Outgoing Heat Flow of Giant Planets on Obliquity and Eccentricity}\label{eosec}

A non-synchronous, fast-rotating planet would naturally homogenize the heat across the longitude toward the zonal-equilibrium state. The previous discussion only considered planets where the incident stellar flux maximizes at the equator by assuming a zero planetary obliquity. We have also assumed the incident stellar flux is constant with time for a planet in a circular orbit. In reality, the rotation axis tilt and the orbital eccentricity often vary (\citealt{ohnoAtmospheresNonsynchronizedEccentrictilted2019,  ohnoAtmospheresNonsynchronizedEccentrictilted2019a}), causing diurnal and seasonal cycles. Studies have been done to understand the effects of planetary eccentricity and obliquity on the atmospheric dynamics, energy balance, and climate of giant planets (e.g., \citealt{langtonObservationalConsequencesHydrodynamic2007,  langtonHydrodynamicSimulationsUnevenly2008, lewisAtmosphericCirculationEccentric2010, katariaThreedimensionalAtmosphericCirculation2013, lewisAtmosphericCirculationEccentric2014,  rauscherModelsWarmJupiter2017, ohnoAtmospheresNonsynchronizedEccentrictilted2019,  ohnoAtmospheresNonsynchronizedEccentrictilted2019a, mayorgaVariableIrradiation1D2021, tanWeakSeasonalityTemperate2022}). But there has not been a study on how the internal heat flow of giant planets is affected by the seasonal change of the incident stellar flux. 

The theory of the inhomogeneity effect suggests that spatial and temporal variability of the stellar flux in the radiative zone should impact the internal heat flow. In this section, we extend previous discussions on spatial inhomogeneity to include the effects of temporal variation. We demonstrate that the total heat flow of a planet can vary depending on its obliquity and eccentricity, even when the total amount of stellar energy received over a year is the same.

\begin{figure}
  \centering \includegraphics[width=0.45\textwidth]{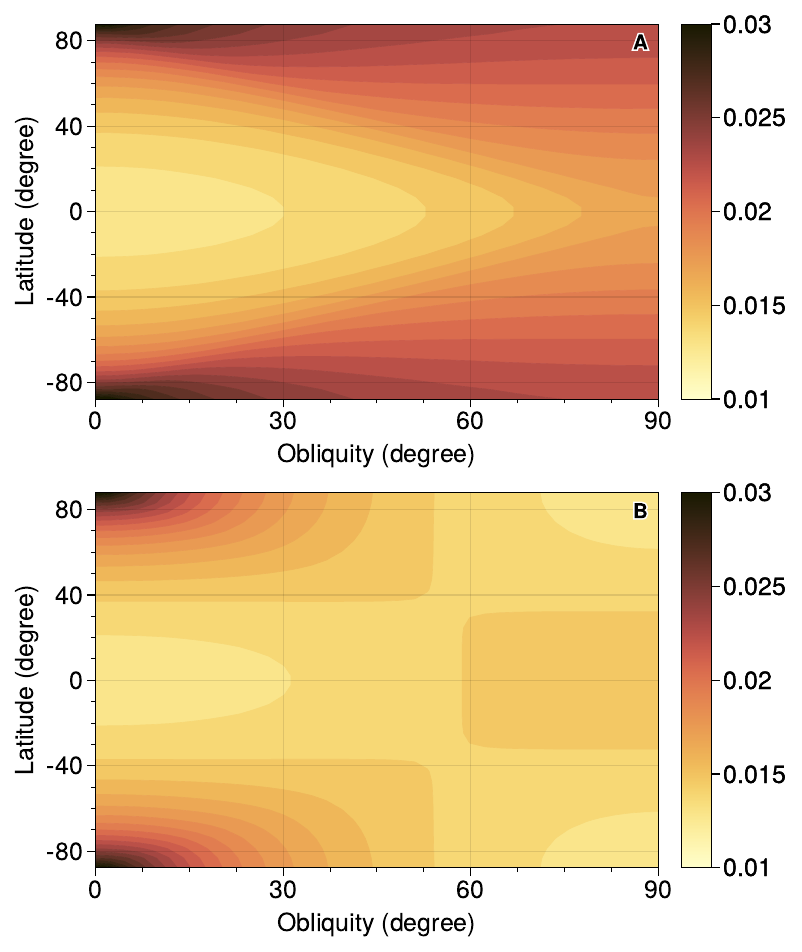} 
  \caption{The orbital-mean internal heat flux for a fast-rotating tilted planet as functions of obliquity and latitude. We set $\beta\sim 0.76$, $D=2$, $K=0.2$, and $\alpha=1$. (A): planets with short radiative timescales so that the incoming stellar flux is homogenized over longitude within each rotational period (diurnal-mean scenario). (B): planets with long radiative timescales so that the incoming stellar flux is homogenized over longitude for the entire year (annual-mean scenario). We assumed no heat transfer across the latitude.} \label{eocontour}
\end{figure}

The orbital eccentricity mainly induces the time variability of the incident stellar flux during its orbit. Planetary obliquity could create a more complex pattern of incident stellar flux distribution on the planet that changes over time. The cosine of the instantaneous stellar incident angle, $\mu$, is (e.g., \citealt{williamsHabitablePlanetsHigh1997}):
\begin{equation}
\mu= \sin \phi \sin \phi_{\rm ss} + \cos \phi \cos \phi_{\rm ss} \sin D_h,
\end{equation}
where $\phi$ is latitude and $\phi_{\rm ss}$ is the substellar latitude. $D_h$ is the hour angle between sunset and sunrise, given by
\begin{equation}
\cos D_h = \max\{\min\{-\tan\phi \tan\phi_{\rm ss}, -1\},-1\}.
\end{equation}
The $\max$ and $\min$ ensure $0<D_h<\pi$ and account for the polar day and night. $\phi_{\rm ss}$ depends on the planetary obliquity $\theta$ and the orbital configuration (see Equation 15 in \citealt{ohnoAtmospheresNonsynchronizedEccentrictilted2019}). 

We will continue using the previous analytical framework and assume that the deep convective layer is constant in space and time due to its efficient dynamics and long radiation timescale. We assume the planetary rotation is fast enough to average the weather within a day. For each day, the stellar flux heating terms ($F_{\rm v}$, $\partial F_{\rm v}/\partial \tau$, and $S$) related to the stellar flux contribution in the RCE solutions are integrated over the hour angle from $-D_h$ to $D_h$ and then averaged over the planet's latitudinal circle. 

We will first investigate the obliquity effect with a fixed eccentricity to zero. The planet in a circular orbit has a constant angular velocity $\omega=2\pi/P$ and $P$ is the orbital period, such that $\sin\phi_{\rm ss}=\sin\theta\cos\omega t$. $\phi_{\rm ss}=0$ at the equinox and $\phi_{\rm ss}=\pm \theta$ at the solstices. If $\theta=0$, $\mu$ goes back to $\mu_\phi=\cos\phi\cos D_h$, consistent with the angular distribution in the LE case. 

We will consider two scenarios. In the first scenario, where the atmosphere has a short radiation timescale, we assume that the atmosphere is in equilibrium for each day. We obtain the zonal-equilibrium solution in RCE for each latitude on each day. This is the diurnal-mean scenario, which allows us to see the effect of a time-varying incident stellar flux pattern during the orbit. In the second scenario, where the radiation timescale is very long (e.g., longer than a planetary year), we will consider an annual-mean scenario by further averaging the heating terms over the entire year at each latitude. The incident stellar flux pattern in the annual-mean scenario varies with obliquity but not time as it is averaged over the year. Both the diurnal-mean and annual-mean scenarios have the same total amount of stellar flux received over a year. However, their different temporal and spatial distributions would lead to different planetary heat flow at the top of the atmosphere. This is the essence of the inhomogeneity effect. 

We used the same parameters as in the ZE case in Section \ref{secrce}. The global-mean outgoing internal heat flux for diurnal-mean and annual-mean incident stellar flux scenarios is averaged over a year and presented in \Cref{eocontour}A and \Cref{eocontour}B, respectively. The diurnal-mean scenario (\Cref{eocontour}A) shows that for low-obliquity planets, a majority of internal heat escapes from high latitudes, consistent with the ``radiator fin" scenario discussed in Section \ref{secrce}. However, due to the smaller surface area of high latitudes, this contribution is small. As obliquity increases, internal heat flux decreases in both high and low latitudes, with the higher-latitude heat flux remaining higher than the lower-latitudes but the latitudinal contrast reducing.

\begin{figure}
  \centering \includegraphics[width=0.45\textwidth]{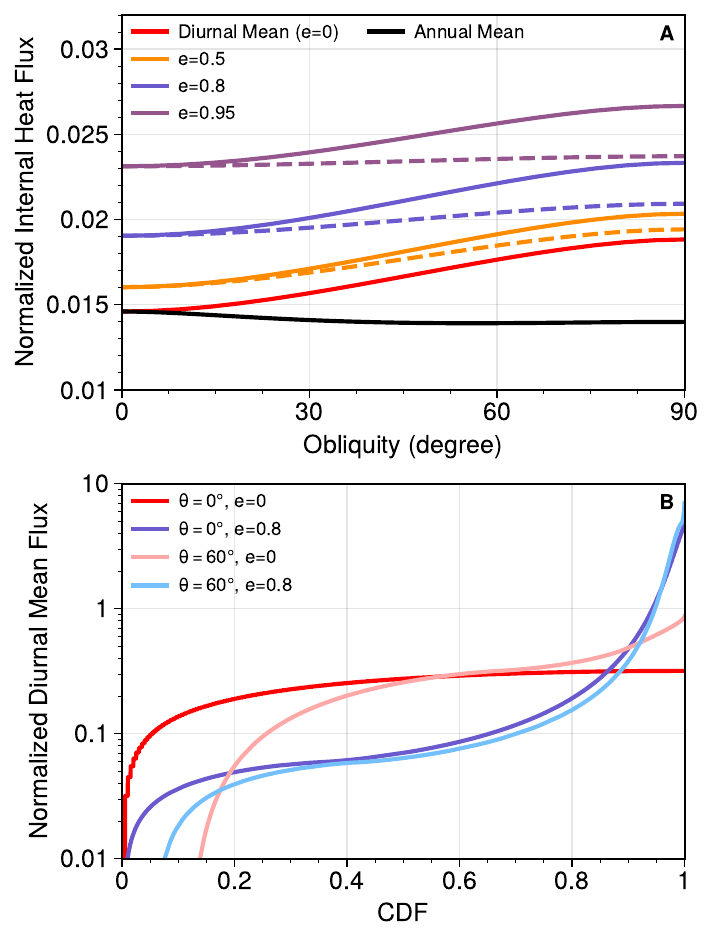} 
  \caption{(A): The normalized orbital-mean global-mean internal heat flux as functions of planetary obliquity and orbital eccentricity. The internal heat flux in the diurnal-mean scenario (color lines) increases with obliquity and eccentricity. The solid lines represent the planets with the summer solstice at the periapse ($f_{\rm eq}=\pi/2$), and the dashed lines represent those with the vernal equinox at the periapse ($f_{\rm eq}=0$). The annual-mean scenario (black) will be further analyzed in \Cref{amficdf}. (B): CDFs of the normalized top-of-the-atmosphere incident stellar flux distribution across the planetary globe and orbit for selected diurnal-mean cases with the vernal equinox at the periapse ($f_{\rm eq}=0$). For a given eccentricity, a larger obliquity yields a larger inhomogeneity of the incident stellar flux (red versus light red, blue versus light blue). Likewise, For a given obliquity, a larger eccentricity yields a larger inhomogeneity of the incident stellar flux (red versus blue). For obliquity of $\theta=60^{\circ}$, the light red curve ($e=0$) and the light blue curve ($e=0.8$) intersect twice such that our theory is not strictly applicable.} \label{eoficdf}
\end{figure}

The global-mean internal heat flux in the diurnal-mean case increases with planetary obliquity (see the color lines in \Cref{eoficdf}A). This behavior is consistent with our theory. Using the equal-area grid, we calculated the CDFs of incident stellar flux distribution for two planets in a circular orbit (\Cref{eoficdf}B). The lower-obliquity planet in the circular orbit (red) intersects once with its counterpart with a higher obliquity (blue). From our metric of inhomogeneity in Section \ref{sec:theory}, the higher-obliquity case exhibits a greater inhomogeneity. Because the internal heat flux is a convex function of incident stellar flux (\Cref{insofig}C), the outgoing internal heat flux should increase with planetary obliquity. On a higher-obliquity planet, the seasonal change in incident stellar flux pattern has a greater temporal variability at each latitude, which in turn significantly increases the total internal heat flux over a year. The internal heat flux increases by approximately 30\% from $\theta=0^{\circ}$ to $90^{\circ}$ in the diurnal-mean case (\Cref{eoficdf}A).

In the annual-mean scenario, the behavior of low-obliquity planets is similar to that in the diurnal-mean scenario, with a large internal heat flux escaping from the high latitudes (\Cref{eocontour}B). However, the latitudinal dependence of obliquity differs in the annual-mean case. As obliquity increases, internal heat flux decreases in the high latitudes but increases in the low latitudes. At $\theta\sim 54^{\circ}$, the heat flux is roughly uniform with latitude because the annual-mean incident stellar flux at higher latitudes is equal to that at lower latitudes (\citealt{ohnoAtmospheresNonsynchronizedEccentrictilted2019}). As a result, latitudinal contrast in the outgoing heat flux decreases with increased obliquity from $0^{\circ}$ to $54^{\circ}$ and increases from obliquity $54^{\circ}$ to $90^{\circ}$.

\begin{figure}
  \centering \includegraphics[width=0.45\textwidth]{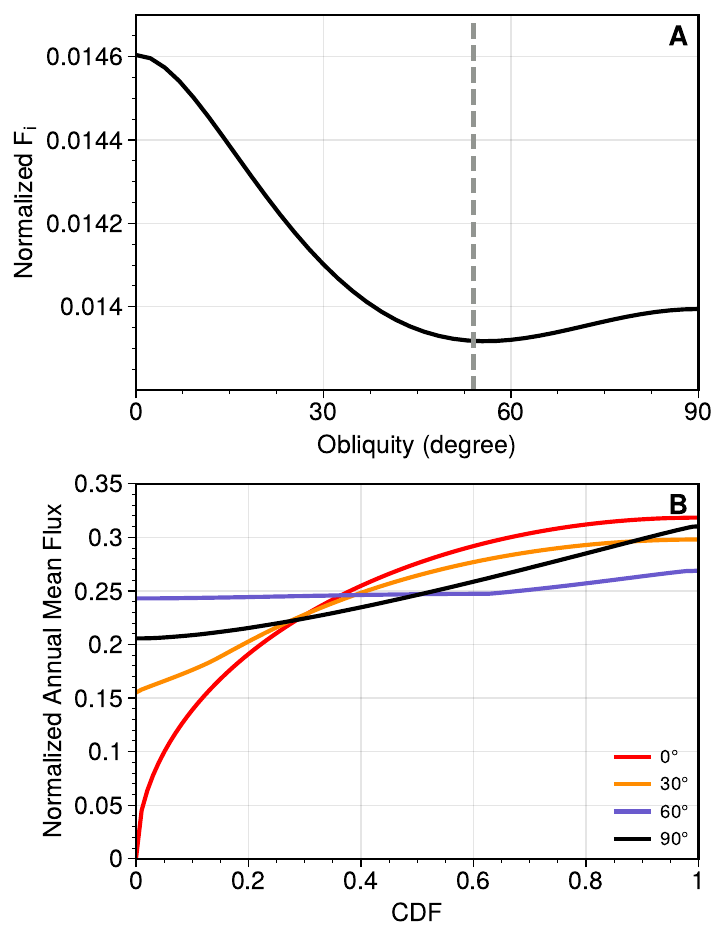} 
  \caption{(A): The Normalized global-mean internal heat flux as a function of obliquity in the annual-mean scenario. The internal heat flux decreases slightly but nonlinearly with obliquity. The minimum flux occurs at around $54^{\circ}$ (dashed line). (B): CDFs of the normalized top-of-the-atmosphere stellar flux distribution across the globe for selected annual-mean cases. The CDFs of $\theta=0^{\circ}$, $30^{\circ}$, and $60^{\circ}$ only intersect once. The inhomogeneity of the incident stellar flux and internal heat flux increase with obliquity when $\theta<54^{\circ}$. But the CDF of the $90^{\circ}$ case appears steeper than the $60^{\circ}$ case, and the outgoing internal flux is higher. The CDFs of the $30^{\circ}$ and $90^{\circ}$ intersect twice such that our theory may not apply in their comparison.} \label{amficdf}
\end{figure}

The pattern of the internal heat flux map in the annual-mean scenario mirrors the incident stellar flux map (see Figure 2 of \citealt{ohnoAtmospheresNonsynchronizedEccentrictilted2019}), showing a decreasing trend in latitudinal inhomogeneity with obliquity when $\theta<54^{\circ}$ and an increasing trend when $\theta>54^{\circ}$. This nonlinear trend is further supported by the CDFs of the incident stellar flux distribution (\Cref{amficdf}B). Inhomogeneity decreases from $\theta=0^{\circ}$ to $30^{\circ}$ and $60^{\circ}$, with the smallest inhomogeneity at $\theta=0^{\circ}$ and the largest at $\theta=90^{\circ}$. As a result, internal heat flux in the annual-mean case decreases with increasing obliquity from 0 to 54 degrees and increases to $\theta = 90^{\circ}$ (\Cref{amficdf}A). However, overall, internal heat flow does not vary significantly with obliquity compared to the diurnal-mean scenario (\Cref{eoficdf}A).

The relationship between total planetary heat flow and obliquity suggests that planets with different obliquities may have different evolutionary paths. A high-obliquity body may cool faster over time. The importance of radiative timescale is highlighted by the two limiting cases in the diurnal-mean and annual-mean scenarios. In the short-timescale limit, internal heat flux increases with obliquity, but in the long-timescale limit, heat flux slightly decreases and exhibits nonlinear behavior. For example, in a planet with $\theta=90^{\circ}$ in the diurnal-mean case, when the star is over one pole, the other hemisphere is in permanent darkness. This is similar to the LE case, and the internal heat flux escapes preferentially from the nightside (polar night in this case). However, if the radiative timescale is long, the annual-mean case develops a two-polar stellar flux pattern with an average stellar flux (\Cref{eocontour}), and the cooling efficiency is greatly reduced. Therefore, internal heat flux can vary significantly depending on the radiative timescale of the atmosphere. 

Next, we explore the effect of eccentricity. Given an orbital eccentricity $e$, the time-varying stellar flux is a function of the true anomaly $f$:
\begin{equation}
F_{\rm v}\propto\big(\frac{1-e^2}{1+e\cos f}\big)^{-2}(1-e^2)^{\frac{1}{2}}.
\end{equation}

We have included a scaling factor $(1-e^2)^{\frac{1}{2}}$ on the right-hand side based on Kepler's law to account for the variation in orbital-averaged stellar flux with eccentricity. This ensures the total stellar flux remains constant in one orbit for different eccentricity. The relationship between the true anomaly $f$ and the eccentric anomaly $E$ is:
\begin{equation}
\tan \frac{f}{2} = \sqrt{\frac{1+e}{1-e}}\tan\frac{E}{2}.
\end{equation}
Kepler's equation of $E$ is written as (\citealt{murraySolarSystemDynamics1999}):
\begin{equation}
\frac{dE}{dt} =\frac{\omega}{1-e\cos E}.
\end{equation}

The substellar latitude of a tilted planet in an eccentric orbit changes with the true anomaly $f$ as:
\begin{equation}
\sin\phi_{\rm ss}=\sin\theta\sin(f-f_{\rm eq}),
\end{equation}
where $f_{\rm eq}$ is the true anomaly of the vernal equinox. At the periapse ($f=0$), the planet is at the vernal equinox when $f_{\rm eq}=0$ and is at the summer solstice when $f_{\rm eq}=\pi/2$. This work considers both $f_{\rm eq}$ cases.

We only need to consider the diurnal-mean cases because the annual-mean stellar flux at each latitude should not change with eccentricity. The results with different eccentricity and obliquity are shown in \Cref{eoficdf}A. Similar to the circular orbit, the internal heat flux emitted by the planet in an eccentric orbit increases as the obliquity increases. Due to the increased time variability of incident stellar flux caused by eccentricity, the planet in an eccentric orbit emits more internal heat flux than in a circular orbit. For extremely eccentric planets such as $e=0.95$, the internal heat flux could be 70\% greater than planets in circular orbits with our adopted parameters. 

This increase in internal heat flux with eccentricity is consistent with the theory of the homogeneity effect. The CDFs of incident stellar flux distribution for eccentric planets are plotted in \Cref{eoficdf}B. For zero-obliquity planets, the CDF of the eccentric planet is more inhomogeneous than its circular counterpart. For obliquity of $\theta=60^{\circ}$, the low- and high- eccentricity cases intersect twice, so our CDF theory is not applicable. One can also compare the cases with different obliquity and eccentricity. For example, the case with $\theta=60^{\circ}$ and $e=0.8$ appears steeper than the planet with $\theta=0^{\circ}$ and $e=0$. 

Varying the true anomaly of the vernal equinox also has a significant impact on the internal heat flux (\Cref{eoficdf}A). The planet with the summer solstice at the periapse ($f_{\rm eq}=\pi/2$) emits more internal heat flux than those with the vernal equinox at the periapse ($f_{\rm eq}=0$). The dependence on the obliquity is also steeper in the former case, indicating a more dramatic difference at higher obliquities. The dependence of $F_{\rm int}$ on $f_{\rm eq}$ attributes to the orbital geometry. We can demonstrate this considering two highly tilted planets with an obliquity of $\theta=90^{\circ}$, one with $f_{\rm eq}=0$ and the other with $f_{\rm eq}=\pi/2$. In a highly eccentric orbit, the most efficient planetary cooling should occur around the apsis ($f=\pi/2$) rather than the periapse ($f=0$) because intense stellar flux at proximity to the star (periapse) suppresses internal heat flux emission and the planet spends more time around the apsis due to Kepler's second law. For the planet with $f_{\rm eq}=\pi/2$, one pole of the planet faces the star at the apsis, similar to the LE case. However, the planet with $f_{\rm eq}=0$ experiences the fall equinox at the apsis, similar to the ZE case. As a result, the planet with $f_{\rm eq}=\pi/2$ emits much more internal heat flux than that with $f_{\rm eq}=0$ at the apsis where most cooling occurs (\Cref{eoficdf}A).

In conclusion, given the same orbital-averaged stellar flux, the specific configuration of the star-planet system can affect the spatial and temporal inhomogeneity of incident stellar flux, which in turn affects the planetary cooling flux at the top of the atmosphere. Eccentricity, obliquity, and the true anomaly of the vernal equinox are key parameters that can lead to significant inhomogeneity and should be considered in future planetary evolution calculations. The radiative timescale of the radiative zone and the heat redistribution timescale across the globe are also important factors. However, in this study, we only considered a short radiative timescale case (``diurnal mean") and a long radiative timescale case (``annual mean") in our RCE framework. Further research with a more detailed treatment of dynamics will provide more insight.

\section{Conclusion and Discussion} \label{consec}

In this study, we generalized the theory of the inhomogeneity effect in Paper I to investigate the impact of planetary parameters on cooling. We introduced the cumulative distribution function to measure the degree of inhomogeneity, allowing comparison among different inhomogeneous planets. We then systematically examined the effect of previously overlooked parameters such as rotation rate, thermal inertia, heat redistribution efficiency, obliquity, and orbital eccentricity on energy redistribution on 3D planets and planetary cooling using analytical models. Key points to note include:  

1. Slow-rotating or tidally locked planets experience larger cooling than fast-rotating planets, which in turn experience larger cooling than the globally-equilibrium scenario like in traditional 0D or 1D models. This applies to both airless bodies and giant planets.

2. The mean surface temperature of airless planets increases with rotational rate and surface thermal inertia, with the tidally locked configuration and equilibrium temperature providing the lower and upper bounds, respectively.

3. On giant planets, internal heat flux escapes preferentially from areas with less incident stellar flux, such as high latitudes or night side, acting as ``radiator fins'' on the planet.

4. The effect of inhomogeneous incident stellar flux is greater when stellar flux can penetrate deeper into the atmosphere, which can occur with colder interiors or smaller visible-to-IR opacity ratios in the radiative zone.

5. If the vertical heat transport on giant planets is governed by radiation, a horizontally conservative heat redistribution does not alter the global-mean internal flow. However, when convection occurs in deep atmospheres, the inhomogeneity effect results in a different global-mean internal flow, which decreases as the efficiency of zonal heat redistribution between day and night sides increases through atmospheric dynamics or planetary rotation.

6. On rapidly rotating planets, the outgoing internal flux generally increases with planetary obliquity and orbital eccentricity. The radiative timescale and true anomaly of the vernal equinox also play significant roles. If the radiative timescale is long, the outgoing internal flux may exhibit a slightly decreasing but nonlinear trend with obliquity.

The caveats of the analytical models have been discussed in Paper I. Here, we mostly focus on the implications of this work. Considering the inhomogeneity effect could bring new insights into the current understanding of planetary evolution. In the traditional 1D framework, an unknown mechanism is needed to explain the inflated sizes of hot Jupiters. Typically, 1D models (e.g., \citealt{thorngrenBayesianAnalysisHotJupiter2018, thorngrenIntrinsicTemperatureRadiative2019,sarkisEvidenceThreeMechanisms2021}) estimated that 2-3\% of incoming stellar flux needs to be deposited below the radiative-convective boundary to offset the cooling from the interior of the most inflated planets. However, with significant atmospheric inhomogeneity, such as in the local-equilibrium case with inefficient heat redistribution between the dayside and nightsides, the cooling flux could be much larger than that predicted by the global-equilibrium model (\Cref{akcontour}). This suggests that the required energy flux may have been underestimated in previous studies. The day-night temperature contrast is generally large for hotter giant planets such as those exceeding 2000 K (e.g., \citealt{komacekAtmosphericCirculationHot2016, zhangAtmosphericRegimesTrends2020}). Our work suggests that the estimated heating efficiencies for hot planets could be higher than currently thought.

Moreover, in Paper I, we found that the opacity inhomogeneity might have an even larger impact on the interior cooling of giant planets than the inhomogeneity of the incident stellar flux. Limited by the analytical framework, we could not explore the 3D distribution of opacity on giant planets and their impact on the cooling flux in this study. In the next paper (\citealt{zhangInhomogeneityEffectIII2023}), we will use a non-hydrostatic general circulation model to examine the inhomogeneity effects of incident stellar flux and opacity, as well as that of atmospheric dynamics in a self-consistent fashion.

For hot Jupiters in eccentric orbits, the interior cooling could be more efficient, and more energy is needed to inflate them. If the planet has undergone high-eccentricity migration (\citealt{dawsonOriginsHotJupiters2018}), the cooling enhancement is notable compared with that in the circular orbit. But high eccentricity might also produce a large tidal heating that could greatly inflate the planet (e.g., \citealt{millerInflatingDeflatingHot2009}).  

Planets with high obliquities could cool faster than the low-obliquity planets if the radiative timescale is not long. One interesting example is Uranus, which emits almost zero internal heat flux in the current state (\citealt{pearlAlbedoEffectiveTemperature1990}). If Uranus' high obliquity leads to faster planet cooling, it might be an important factor in the Uranus-Neptune dichotomy---twin planets with similar mass and size but different internal heat fluxes. For the current Uranus, although the latest calculation suggested that the radiative timescale in the upper troposphere and stratosphere is a few tens of years, smaller than its orbital timescale (\citealt{liHighperformanceAtmosphericRadiation2018}), observations found no detectable time variability of the meridional distribution of temperature over a season (\citealt{ortonThermalImagingUranus2015,romanUranusNorthernMidspring2020}), implying a longer radiative timescale than its orbital timescale. From our analysis above, the high obliquity might not help accelerate the interior cooling of the current Uranus compared with Neptune. However, Uranus was hotter before its internal heat flux was gone when its radiative timescale was short (e.g., see the Neptune case in \citealt{liHighperformanceAtmosphericRadiation2018}). In this scenario, higher obliquity would lead to faster cooling compared to Neptune. It is important to remember that the effects of inhomogeneity are not very pronounced when the internal heat flux is substantial (e.g., a large $K$). Given that Uranus and Neptune receive weak irradiation, the impact of inhomogeneity is likely negligible during their early hot stage. This might be also true for present-day Neptune, where the internal heat flux exceeds the outgoing heat flux.

Another important factor is that the high obliquity leads to significant variation in the photochemical species in the stratosphere of Uranus (\citealt{mosesSeasonalStratosphericPhotochemistry2018}). Uranus (and other giant planets in the solar system) have significant cloudy dynamical processes where the moist convection transports a larger amount of internal energy out. Cloud activity on Uranus appeared to increase as the planet approached the 2007 equinox (e.g., \citealt{huesoAtmosphericDynamicsVertical2019,fletcherIceGiantSystems2020}). The time variability in atmospheric opacity from gas and clouds might produce large atmospheric inhomogeneity to affect interior cooling that we did not discuss in this study. Whether the inhomogeneity effect plays a significant role in the Uranus-Neptune dichotomy deserves more diagnosis in the future. 

The inhomogeneity effect also raises a question about properly averaging the stellar flux in 1D models of irradiated planets. Traditional 1D models (e.g., \citealt{fortneyPlanetaryRadiiFive2007, guillotRadiativeEquilibriumIrradiated2010}) used global-mean stellar fluxes (i.e., $1/4$) but varied the incident angle (or $\mu$) to approximate the globally averaged incident angle. Using the same approach to reproduce the global-mean internal heat flux from our 3D model in the local-equilibrium state, we can estimate the effective $\mu$ in the 1D model. In the cases in \Cref{akcontour}, for $\alpha$ is 0.1, and $K$ is 0.01, the effective $\mu$ is about 0.025 (or the incident angle is about 88.5 degrees). If $\alpha$ is 1 and $K$ is 5, the effective $\mu$ is about 0.7 (or the incident angle is about 45.6 degrees). Therefore, the effective $\mu$ depends on the values of $\alpha$ and $K$. There is no universal effective $\mu$ that can be used in a 1D model to mimic the effect of inhomogeneous stellar flux distribution on a 3D planet.

Finally, the inhomogeneity effect can be validated through several observational implications on the spatial and temporal variabilities of the outgoing internal heat flux of giant planets. Jupiter is almost not tilted, but several long-term oscillatory modes have been observed (\citealt{ortonUnexpectedLongtermVariability2023}). The tilted giant planets should show a time evolution of internal heat flux if the radiative timescale is not long. This is distinct from the seasonal variation of planetary emission flux. High-precision data may be used to detect this variability on Saturn and Neptune, despite potential complications from seasonal changes in stratospheric chemical species. 

It is important to note that our simplified RCE column model does not take into account atmospheric dynamics. While the ZE case demonstrates a decrease in internal heat flux from the equator to the pole (\Cref{insofig}F), this change in flux does not fully compensate for the latitudinal variation of incident stellar flux. As a result, the ZE model is expected to exhibit significant equator-to-pole variations in total outgoing thermal emission at the top of the atmosphere. This finding contrasts with the nearly uniform emission flux observed in Jupiter (e.g., \citealt{ingersollPioneer11Infrared1975}). Several hypotheses have been proposed to explain the uniformity of the emission flux, including mixing in the upper cloud layers (e.g., \citealt{conrathGlobalVariationPara1984,pirragliaMeridionalEnergyBalance1984}), a latitudinal temperature gradient in the deep atmosphere (\citealt{ingersollPioneer10111976,ingersollSolarHeatingInternal1978}), or a stronger polar convective heat flux compared to the equator based on 3D interior simulations (e.g., \citealt{aurnouConvectiveHeatTransfer2008}). To provide a quantitative explanation for real-world data, a dynamical model of giant planets is required. In the next paper (\citealt{zhangInhomogeneityEffectIII2023}), we will present an example that investigates the impact of general circulation on heat transport in tidally locked exoplanets.

Another implication is that tidally locked giant exoplanets should have internal heat flux primarily emanating from the night side rather than the dayside due to the intense incident stellar flux on the dayside that reduces the internal energy flux. Although the dayside emission is larger and much easier to be detected than the nightside flux, the small signal of the internal IR flux could be significantly contaminated by the re-emission of the stellar energy on the dayside. Detecting the nightside infrared excess during a primary eclipse can help to confirm that the internal heat flux is mainly emitted at the nightside on these tidally locked giants. Additionally, among planets with similar parameters, those with a larger day-night contrast will emit larger internal heat fluxes. With a large sample in the JWST era, future population analysis may be able to test this prediction.

\section{acknowledgments}

We acknowledge the reviewer for their careful reading and helpful comments. X.Z. is supported by NASA Exoplanet Research Grant 80NSSC22K0236 and NASA Interdisciplinary Consortia for Astrobiology Research (ICAR) grant 80NSSC21K0597. This work was inspired by the last in-person conversation with Dr. Adam P. Showman (1969-2020) about the heat flow on tidally locked exoplanets at the AGU Fall Meeting 2019. We also acknowledge the assistance provided by the OpenAI language model.

%

\vspace{5mm}

\appendix
\section{dependence of the convex function mean on inhomogeneity}
\label{app:convex}

First, let us prove a theorem for two sets of discrete variables.

\textit{Theorem I.} For any convex function $f(x)$ and two sets of random variables $X = \{x_1, x_2, ..., x_n\}$ and $Y = \{y_1, y_2, ..., y_n\}$ with the same expected value $\overline{X} = \overline{Y}$. Without loss of the generality, we assume $x_1 \le x_2 \le, ..., \le x_n$ and $y_1 \le y_2 \le, ..., \le y_n$. If the differential sequence $x_i-y_i$ has one sign change from negative to positive as $i$ increases from 1 to $n$, i.e., there exists $y_c$ in $Y$ such that $(y_i-y_c)(x_i-y_i) \ge 0$ for all $i$, then
\begin{equation}
\overline {f(X)} \ge \overline {f(Y)}. 
\end{equation}

\textit{Proof.} Since $f(x)$ is convex,
\begin{equation}
f(x_i)-f(y_i) \ge f^\prime(y_i)(x_i-y_i), \label{app2:c1}
\end{equation}
where $f^\prime(y_i)$ is the derivative of $f$. Also, the convexity of $f$ implies 
\begin{equation}
\frac{f^\prime(y_i)-f^\prime(y_c)}{y_i-y_c} \ge 0. \label{app2:c2}
\end{equation}
Using $(y_i-y_c)(x_i-y_i) \ge 0$, we obtain
\begin{equation}
(x_i-y_i)\left[f^\prime(y_i)-f^\prime(y_c)\right] \ge 0. \label{app2:c3}
\end{equation}
Add \Cref{app2:c1} and \Cref{app2:c3} to get:
\begin{equation}
f(x_i)-f(y_i) \ge f^\prime(y_c)(x_i-y_i).
\end{equation}
Then sum over all $i$ on both sides and take the average. Because $\overline{X}=\overline{Y}$ and $f^\prime(y_c)$ is a constant, the right-hand side vanishes. We find $\overline {f(X)-f(Y)} \ge 0$ and prove the theorem
\textit{Theorem I.}

We have compared the ``inhomogeneity" between two sets of variables using the expression $(y_i-y_c)(x_i-y_i)$. If this expression is strictly positive for two monotonically increasing sequences, set X has a larger contrast than set Y, and vice versa. It is clear that the condition $(y_i-y_c)(x_i-y_i) \ge 0$ implies that the variance of X is larger than that of Y, because the variance, $\mathrm{var}(X)\equiv\overline{(X-\overline{X})^2}$, is a convex function of $X$. However, we claim that $(y_i-y_c)(x_i-y_i) \ge 0$ is stronger than just comparing variances. Proof of this can be provided.

\textit{Proposition:} If $(y_i-y_c)(x_i-y_i) \ge 0$ for all $i$, then their variances satisfy:
\begin{equation}
\mathrm{var}(X)-\mathrm{var}(Y) \ge \overline{(X-Y)^2}. 
\end{equation}

\textit{Proof.} The condition implies:
\begin{equation}
(y_i-y_c)(x_i-y_i)=\frac{1}{2}(x_i+y_i-x_i+y_i-2y_c)(x_i-y_i) \ge 0.
\end{equation}
Thus
\begin{equation}
(x_i^2-y_i^2) \ge (x_i-y_i)^2+2y_c(x_i-y_i).
\end{equation}
Then sum over all $i$ on both sides and take the average. Note that $\overline{X}=\overline{Y}$ and $y_c$ is a constant:
\begin{equation}
\overline{X^2-Y^2} \ge \overline{(X-Y)^2}.
\end{equation}
Because the variance of $X$ satisfies $\mathrm{var}(X)\equiv\overline{(X-\overline{X})^2} =\overline{X^2-\overline{X}^2}$ and $\overline{X}=\overline{Y}$, we have
\begin{equation}
\mathrm{var}(X)-\mathrm{var}(Y) \ge \overline{(X-Y)^2} \ge 0.
\end{equation}

The difference between the two variances is greater than the quadratic difference between the two sequences. Additionally, $(y_i-y_c)(x_i-y_i) \ge 0$ is a more specific requirement for each $i$, and it necessitates only one sign change in the domain, making it stronger than just comparing variances.

For continuous random variables described by probability density distributions (PDF), the condition $(y_i-y_c)(x_i-y_i) \ge 0$ can be replaced using cumulative distribution functions (CDF). The following theorem can be proven.

\textit{Theorem II}. For any convex function $f(x)$ and two PDFs $g(x)$ and $h(x)$ are defined for non-negative numbers in $[0,\infty]$. The two distributions yield the same expected value of $x$, i.e., $\overline{x} := \int_0^{\infty} xg(x) dx = \int_0^{\infty} xh(x) dx$.
Define $G(x)$ and $H(x)$ are the cumulative density distributions (CDF) of $g(x)$ and $h(x)$, respectively. If the function $G(x)-H(x)$ has one sign change from positive to negative as $x$ increases, i.e., there exists $x_c \in [0,\infty]$ such that $(x-x_c)\left[G(x)-H(x)\right] \le 0$, then the expected values of the function $f(x)$ on two distributions satisfy:
\begin{equation}
\int_0^{\infty} f(x)g(x) dx \ge \int_0^{\infty} f(x)h(x) dx. 
\end{equation} 

\textit{Proof.} Because $G^\prime(x)=g(x)$ and $H^\prime(x)=h(x)$, we have
\begin{equation}
\begin{split}
\int_0^{\infty} \left[f(x)g(x)-f(x)h(x)\right] dx &= \int_0^{\infty} f(x) dG(x) - \int_0^{\infty} f(x) dH(x) \\
&= f(x)\left[G(x)-H(x)\right] \Big |_0^{\infty} - \int_0^{\infty} f^\prime (x)\left[G(x)-H(x)\right] dx\\
&= - \int_0^{\infty} f^\prime (x)\left[G(x)-H(x)\right] dx. \label{con:c1}
\end{split} 
\end{equation} 
In the last step, we have used the property of the CDF: $G(0)=H(0)=0$ and $G(\infty)=H(\infty)=1$. If we let $f(x)=x$, the difference between the expected values of $x$ for the two PDFs can be obtained, which is zero.
\begin{equation}
0=\int_0^{\infty} \left[xg(x)-xh(x)\right] dx =- \int_0^{\infty} \left[G(x)-H(x)\right] dx. \label{con:c2}
\end{equation} 
Since $f(x)$ is convex,
\begin{equation}
\frac{f^\prime(x)-f^\prime(x_c)}{x-x_c} \ge 0. 
\end{equation}
Because $(x-x_c)\left[G(x)-H(x)\right] \le 0$, we have
\begin{equation}
\left[f^\prime(x)-f^\prime(x_c)\right]\left[G(x)-H(x)\right] \le 0. 
\end{equation}
Thus, \Cref{con:c1} becomes: 
\begin{equation}
\int_0^{\infty} \left[f(x)g(x)-f(x)h(x)\right] dx \ge \int_0^{\infty} f^\prime (x_c)\left[G(x)-H(x)\right] dx.
\end{equation} 
Because $f^\prime (x_c)$ is a constant, using \Cref{con:c2}, we proved the theorem.
\begin{equation}
\int_0^{\infty} \left[f(x)g(x)-f(x)h(x)\right] dx \ge 0.
\end{equation}

\section{A simple analytical estimate of radiative-convective boundary and internal heat flux}
\label{app:anarcb}

In this section, we derive analytical expressions for the RCB and internal heat flux, using certain approximations. This approach allows for a more intuitive physical understanding compared to the semi-analytical radiative-convective equilibrium (RCE) model used in this study.

In Paper I, the RCE model was established based on two constraints to determine the RCB:

(1) Temperature continuity (Equation 21 in Paper I):  
\begin{equation}
T_{\rm rad}(\tau_{\rm rcb})=T_{\rm conv}(\tau_{\rm rcb}).\label{tcon}
\end{equation}

(2) Upward radiative flux continuity (Equation 22 in Paper I):  
\begin{equation}
F^+_{\rm rad}(\tau_{\rm rcb})=F^+_{\rm conv}(\tau_{\rm rcb}).
\end{equation}
The RE temperature $T_{\rm rad}$ is (Equation \ref{trad}):
 \begin{equation}
    T_{\rm rad}^4(\tau)=S(\tau)+(D+D^2\tau)F_{\rm int}.
\end{equation}
The adiabatic temperature profile in the convective zone $T_{\rm conv}$ is (Equation \ref{tconv}):
\begin{equation}
T_{\rm conv}^4(\tau)=K\tau^\beta. \label{newtconv}
\end{equation}
$F^+_{\rm rad}(\tau_{\rm rcb})$ and $F^+_{\rm conv}(\tau_{\rm rcb})$ are the upward radiative fluxes at the RCB calculated using the radiative-equilibrium temperature $T_{\rm rad}$ and deep adiabatic temperature $T_{\rm conv}$, respectively.

In the first approximation of our simpler analytical model here, we simplified the temperature profile in the radiative zone, $T_{\rm rad}$, by assuming the stellar flux contribution, $S(\tau)$, to be constant. Consequently, a simplified temperature profile can be expressed as:
\begin{equation}
T_{\rm rad}^4(\tau)\sim 2DF_{\odot}+(D+D^2\tau)F_{\rm int}, \label{newtrad}
\end{equation}
where $F_{\odot}$ represents the incoming stellar flux. The approach we used to formulate this simplified radiative-equilibrium temperature profile is different from previous analytical works (e.g., \citealt{bodenheimerRadiiExtrasolarGiant2003,arrasThermalStructureRadius2006,youdinMechanicalGreenhouseBurial2010,ginzburgHotJupiterInflationDue2015}). It can be demonstrated that upon vertical integration (refer to Equation C16 in Appendix B of Paper I), the upward infrared (IR) radiative heat flux at the top of a giant planet's atmosphere is:
\begin{equation}
F_{\rm rad}^+(\tau=0)=\frac{1}{2}\int_0^\infty \left[2DF_{\odot}+(D+D^2\tau)F_{\rm int}\right] e^{-D(t-\tau)}dt
=F_{\odot}+F_{\rm int}.
\end{equation}
Note that the downward IR flux is zero. Hence, with our radiative equilibrium temperature profile in Equation \ref{newtrad}, the incoming stellar flux and the internal heat flux balance the outgoing IR flux.

In our second approximation, we utilized the Schwarzschild criterion rather than the radiative flux continuity constraint to pinpoint the location of the RCB. Although the Schwarzschild criterion may possess certain limitations when estimating the RCB for terrestrial planets (as discussed in Section 3.3 in Paper I), it appears to be a reliable assumption for giant planets, which lack a surface, as used in previous studies of hot Jupiters (e.g., \citealt{ginzburgHotJupiterInflationDue2015}). The Schwarzschild criterion is also considerably simpler than the radiative flux continuity constraint. The criterion states that the gradients of $T_{\rm rad}^4$ and $T_{\rm conv}^4$ are equivalent at the RCB:
\begin{equation}
\frac{dT_{\rm rad}^4}{d\tau}\Big|_{\tau_{\rm rcb}}=\frac{d T_{\rm conv}^4}{d \tau}\Big|_{\tau_{\rm rcb}}.
\end{equation} 
With the new $T_{\rm rad}$ in \Cref{newtrad} and $T_{\rm conv}$ in \Cref{newtconv}, the two constraints that determine the RCB in our simple model are:
\begin{equation}
\begin{split}
    2DF_{\odot}+(D+D^2\tau_{\rm rcb})F_{\rm int} &= K\tau_{\rm rcb}^\beta,\\
    D^2 F_{\rm int} &= \beta K\tau_{\rm rcb}^{\beta-1}.
\end{split}
\end{equation}
One can first eliminate $F_{\rm int}$ to obtain the equation of $\tau_{\rm rcb}$:
\begin{equation}
2D^2F_{\odot}\tau_{\rm rcb}^{1-\beta}=(1-\beta)DK\tau_{\rm rcb}-\beta K. \label{taueq}
\end{equation}
For self-luminous bodies, $F_{\odot}=0$, thus the solutions of $\tau_{\rm rcb}$ and $F_{\rm int}$ are:
\begin{equation}
\begin{split}
\tau_{\rm rcb}&=\frac{\beta}{(1-\beta)D},\\
F_{\rm int}&=\beta^{\beta}(1-\beta)^{1-\beta}D^{-(1+\beta)}K.
\end{split}
\end{equation}
With $D=2$ and $\beta\sim 0.76$, the RCB optical depth, $\tau_{\rm rcb}$, approximates to 1.60, and the internal heat flux, $F_{\rm int}$, is about 0.17$K$. These values are closely aligned with the more precise semi-analytical estimates provided in Section 5.1 of Paper I, where $\tau_{\rm rcb}\sim 1.22$ and $F_{\rm int}\sim0.17 K$.

With high stellar irradiation, the RCB migrates deeper into the atmosphere. If the RCB is sufficiently deep, the first term on the right-hand side of Equation \ref{taueq} should notably exceed the second term ($\beta K$). To derive an explicit form of the $\tau_{\rm rcb}$ solution, we neglect the second term on the right-hand side of Equation \ref{taueq}. The approximate solutions of $\tau_{\rm rcb}$ and $F_{\rm int}$ for irradiated gas giants are:
\begin{equation}
\begin{split}
\tau_{\rm rcb}&\sim\left[\frac{2DF_{\odot}}{(1-\beta)K}\right]^{1/\beta},\\
F_{\rm int}&\sim\frac{\beta K^{1/\beta}}{D^2}\left(\frac{1-\beta}{2DF_{\odot}}\right)^{(1-\beta)/\beta}.
\label{anarcb}
\end{split}
\end{equation}
Given $D=2$ and $\beta\sim 0.76$, the RCB optical depth $\tau_{\rm rcb}\sim 40.57 (F_{\odot}/K)^{1.31}$, and the internal heat flux $F_{\rm int}\sim0.08K^{1.31}F_{\odot}^{-0.31}$. As the incident stellar flux increases, the RCB optical depth also increases but the internal heat flux reduces. Conversely, the RCB moves to the upper atmosphere and the internal heat flux amplifies if the interior is hotter (i.e., when $K$ is larger). For the GE case in this study with $F_{\odot}=1/4$ and $K=0.2$, our estimated $\tau_{\rm rcb}\sim 37.84$ and $F_{\rm int}\sim 0.01471$. The semi-analytical RCE model gives $\tau_{\rm rcb}\sim 68.17$ and $F_{\rm int}\sim 0.01392$. Our straightforward estimation of the RCB optical depth is accurate within a factor of 2, and the internal heat flux appears to be consistent. A detailed comparison of the LE and ZE cases between our simplified and semi-analytical RCE models is depicted in Figure \ref{insofig}.




\end{document}